\documentclass[letterpaper]{article} 
\usepackage{aaai2026}  
\usepackage{times}  
\usepackage{helvet}  
\usepackage{courier}  
\usepackage[hyphens]{url}  
\usepackage{graphicx} 
\usepackage{natbib}  
\usepackage{caption} 
\usepackage[algo2e,vlined,linesnumbered]{algorithm2e}
\usepackage{algorithm}
\usepackage{algorithmic}
\usepackage{pdfpages}
\usepackage{amssymb}
\usepackage{graphicx}
\usepackage{textcomp}
\usepackage{xcolor}
\usepackage{enumerate}
\usepackage{subfigure}
\usepackage{booktabs}
\usepackage{bm}
\usepackage{tcolorbox}
\usepackage{marginnote}
\usepackage{amsmath}
\newcommand{\nop}[1]{}

\newtheorem{definition}{Definition}
\newtheorem{example}{Example}
\newtheorem{lemma}{Lemma}
\newenvironment{proof}
{\par\noindent\textbf{Proof.}\ }
{\hfill$\square$\par}

\usepackage{multirow} 
\usepackage{newfloat}
\usepackage{listings}
\DeclareCaptionStyle{ruled}{labelfont=normalfont,labelsep=colon,strut=off} 
\floatstyle{ruled}
\newfloat{listing}{tb}{lst}{}
\floatname{listing}{Listing}
\title{Effective and Efficient Rumor Assertion Over Large-Scale Social Networks (Technical Report)}
\author {
    Qi Wen\textsuperscript{\rm 1},
    Vipin Chaudhary\textsuperscript{\rm 2},
    Xiang Lian\textsuperscript{\rm 2,3},
    Yutong Ye\textsuperscript{\rm 1,4},
    Mingsong Chen\textsuperscript{\rm 1,4}
}
\affiliations {
    \textsuperscript{\rm 1}East China Normal University,
    \textsuperscript{\rm 2}Case Western Reserve University,
    \textsuperscript{\rm 3}Kent State University,
    \textsuperscript{\rm 4}Zhejiang Sci-Tech University\\
    qiwen7@stu.ecnu.edu.cn, vipin.chaudhary@case.edu, xlian@kent.edu, 52205902007@stu.ecnu.edu.cn, mschen@sei.ecnu.edu.cn
}

\usepackage{bibentry}

\begin{document}

\maketitle

\begin{abstract}

Nowadays, social networks, like Facebook, X, Reddit, YouTube, and so on, have become essential online platforms for users to connect with friends daily, post/share news, and express their opinions/comments. Meanwhile, however, the rapid and widespread dissemination of rumors and misinformation across social networks has become a serious concern. Such misleading information can distort public opinion and potentially cause significant social and economic consequences in real-world scenarios, including product promotion, elections and voting, and even cyberbullying. Therefore, accurately and efficiently verifying whether news spread on social networks is rumor or truth has become an increasingly important yet challenging problem. While previous works usually identified rumors by their contents, in this paper, we propose a novel \textit{user behavior model} to capture the users' judgment/discriminative abilities, which is learned via our well-designed 
evidence-aware probability reconstructor. We develop a general framework for tackling \textit{rumor assertion over large-scale social networks} (RA-SN), which can efficiently and effectively determine whether a given news item is a rumor with high confidence, based on users' behaviors, in a scalable manner. Extensive experiments demonstrate the efficiency and effectiveness of our proposed RA-SN approach on both real-world and synthetic graphs.


\end{abstract}

\section{Introduction}

Recently, social networks have become widely used in people's daily lives to post/share news and opinions with friends, and many real-world applications have emerged, such as online advertising/marketing, knowledge graph discovery, community search/detection, and social network analytics. However, among the vast amount of news spread across social networks, rumors and misinformation may lead to serious consequences such as public panic, reputational damage, and the spread of false beliefs. It is therefore challenging to accurately and efficiently determine whether news is a rumor. 

While prior works~\cite{bian2020rumor, shu2017fake, zubiaga2018detection,qazvinian2011rumor, vosoughi2015automatic, li2019rumor} usually identified content-based rumors, it has become extremely challenging to distinguish rumors based solely on content, owing to the sophisticated design of misinformation and recent advances in generative AI techniques. In contrast, this paper formally defines the problem of \textit{rumor assertion over social networks} (RA-SN) and proposes using human wisdom (i.e., social-network users' discriminative ability) to judge whether a news/message on social networks is a rumor or the truth. 

Below, we provide a motivating example of our RA-SN problem.

\begin{figure}
    \hspace{-4ex}
    \includegraphics[width=1.1\linewidth]{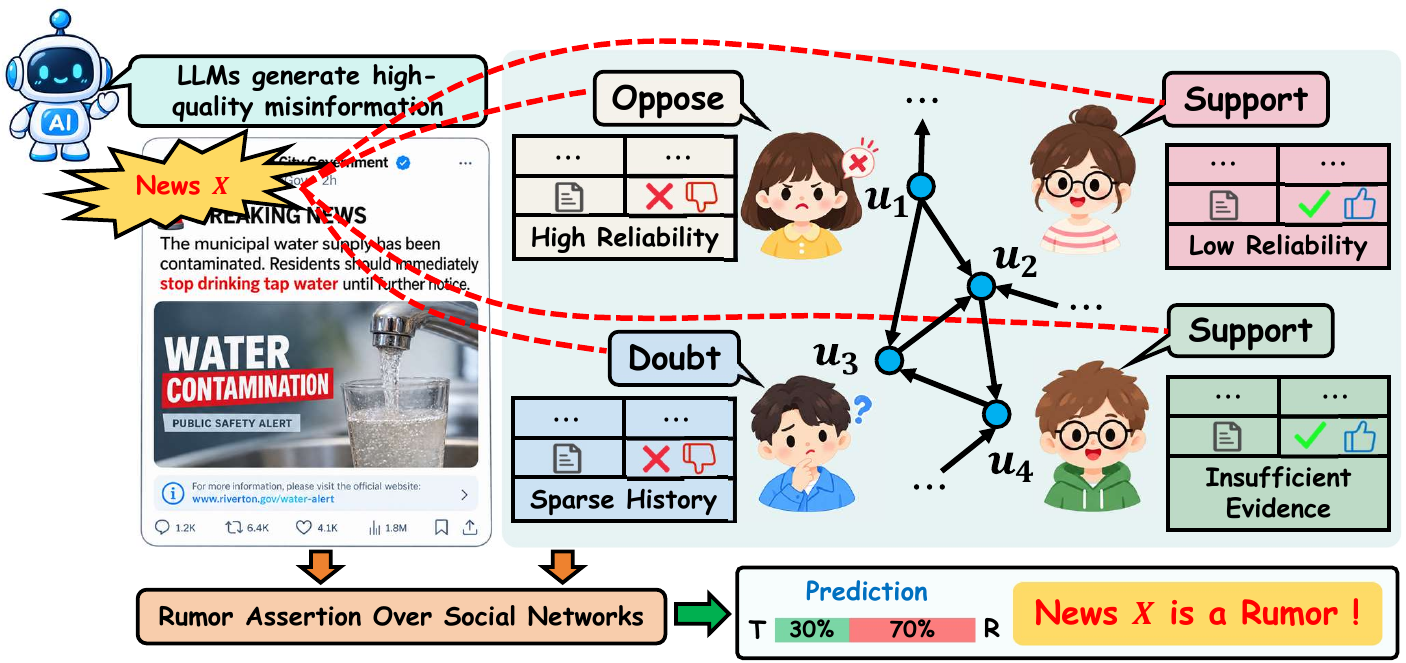}
    \vspace{-2ex}
    \caption{A motivating example of the rumor assertion over social networks (RA-SN).}
    \label{fig:motivation}
    \vspace{-3ex}
\end{figure}

\begin{example} 
{\bf{(Fake News on Social Networks)}} 
Consider a news item $X$ circulating on social networks as shown in Figure~\ref{fig:motivation}, claiming that \textit{``the municipal water supply has been contaminated, and residents should immediately stop drinking tap water''}. Today, Large Language Models (LLMs) can automatically generate such misinformation, accompanied by fluent text, realistic images, and even seemingly official announcements. This makes it increasingly difficult to judge authenticity based solely on the content itself.

Suppose that after news $X$ is published, several users participate in its dissemination.
Some users support or re-post the message, while others raise doubts or express opposition in Figure~\ref{fig:motivation}.
However, these opinions should not all be treated equally.
For example, user $u_1$ has historically supported real news and opposed rumors in most cases, indicating a strong ability to identify misinformation.
In contrast, user $u_2$ frequently forwards or supports false information; thus, the evidence he provides is less reliable.
User $u_3$ has a sparse interaction history, and user $u_4$ has hardly participated in the rumors, making it difficult to assess their discernment capabilities directly.

Therefore, even if most observed users support news $X$ (i.e., 2 supports, 1 opposition, and 1 abstain), it is very unlikely that news $X$ is true (as reliable user $u_1$ reports $X$ as a rumor, while unreliable user $u_2$ comments $X$ as true).
In this case, we can issue our RA-SN query, which uses users' ability to discern genuine news from rumors by analyzing their historical reactions. This way, we can assert that news $X$ is a rumor. \qquad $\blacksquare$


\end{example}

The RA-SN problem has many other real-world applications such as monitoring online public opinion~\cite{ceron2016social}, detecting financial fraud messages~\cite{chang2022design}, fact-checking on social media~\cite{drolsbach2024community}, and verifying AI-generated fake news~\cite{cazzamatta2025ai}.


Solving the RA-SN problem and accurately asserting whether a news item is a rumor with high confidence over large-scale social networks in a scalable manner is challenging. Therefore, in this paper, we propose a novel \textit{user behavior model} to capture users' responses to neighbors' messages, implicitly encoding users' ability to distinguish rumor from truth. We design an effective GNN-based learning technique to model/learn such user behaviors, which can be used for online rumor assertion of a given news item. We also devise an index over message embeddings from social networks to enable fast access to relevant messages for rumor inference.

We list the following major contributions of this paper.

\begin{enumerate}
    \item We formalize the problem of \textit{rumor assertion over social networks} (RA-SN), which determines whether or not a given news item is a rumor.

    \item We propose a novel user behavior model, learned by a GNN-based evidence-aware probability reconstructor, to model how users disseminate news across social networks.

    \item We develop an efficient, scalable algorithm to retrieve (historical) relevant news posted by users and use it for rumor assertion/inference.

    \item We conduct extensive experiments over real-world and synthetic graphs to demonstrate the effectiveness and efficiency of our proposed RA-SN approach under various parameter settings.
\end{enumerate}


\section{Problem Definition}
\label{sec:problem_def}

\setlength{\abovedisplayskip}{0pt}
\setlength{\belowdisplayskip}{0pt}

\subsection{Graph Data Model}
\label{sec:graph_model}

We first provide the data model of a social-network graph, $G$, with rich text information below, which has many real applications~\cite{wasserman1994social}, such as online advertising/marketing, social network analysis, and so on.

\begin{definition} 
\textbf{(Social-Network Graph, $G$)}
A \textit{social-network graph}, $G$, is represented by a triple $(V(G), E(G), \phi(G))$, where $V(G)$ is a set of vertices $v_i$, $E(G)$ is a set of edges $e_{ij}$ ($=e_{v_i, v_j}$) between vertices $v_i$ and $v_j$, and $\phi(G)$ is a mapping function $V(G)$$\times$$V(G)$$\rightarrow$$E(G)$.

Here, each vertex $v_i$ is associated with the posted/forwarded/commented textual messages, denoted as $v_i.M$.



Each edge $e_{ij}$ is associated with an influence score $e_{ij}.inf$, indicating the influence intensity between two friend vertices $v_i$ and $v_j$.
\label{def:graph_model}
\end{definition}



In Definition~\ref{def:graph_model}, a social-network graph $G$ usually contains rich information such as 
short/long texts $v_i.M$ (e.g., tweets, retweets, comments, replies, etc.).
We assume that the directed edges $e_{ij}$ and $e_{ji}$ may have distinct influence scores, $e_{ij}.inf$ and $e_{ji}.inf$, respectively. The influence scores can be inferred by the influence propagation model over social networks, such as \textit{Independent Cascade} (IC) or \textit{Linear Threshold} (LT) model \cite{DBLP:conf/kdd/KempeKT03}.




\subsection{User Behavior Model}


While the information spread on social networks can be either truth or rumor, it is rather challenging to distinguish them simply by their contents only (especially with generative AI maturing nowadays, it can produce contents that appear true, but may actually be misinformation). Therefore, in this paper, instead of the rumor assertion via contents, we aim to distinguish such (mis)information based on users' behaviors/reactions (e.g., posting/forwarding/commenting on messages).

\vspace{0.5ex}\noindent {\bf Truth and Rumor.} We first define truths and rumors on social networks.

\begin{definition} \textbf{(Truth and Rumor)}
Given a message $v_i.M$ from a social-network graph $G$, if message $v_i.M$ is true, we say it is a \textit{truth} (denoted as $T$); otherwise, it is a \textit{rumor} (denoted as $R$).
\label{def:truth_rumor}
\end{definition}

In Definition \ref{def:truth_rumor}, to check whether or not a message $v_i.M$ is true, we can compare message $v_i.M$ (e.g., posts, forwarded messages, or comments) and with groundtruth data from reliable sources (e.g., official websites/accounts, etc.). Message $v_i.M$ is labeled by either truth $T$ or rumor $R$.


\noindent {\bf User Behaviors.} We next formalize the behavior of a user vertex $v_i$, in response to the truth or rumor posted by his/her friends (neighbors) in social networks.

\begin{definition} \textbf{(User Behavior)} Given a graph $G$, assume that a user vertex $v_i$ has a set, $N(v_i)$, of 1-hop neighbor(s) $v_j$. 
User $v_i$ posts a message $v_i.M$, in response to the message $v_j.M$ from one's neighbor $v_j\in N(v_i)$, in the following 6 cases of user response behaviors (as given in Table \ref{tab:user_behavior}):



\begin{itemize}
    \item {\bf Case 1:} When $v_j.M$ is a truth ($T$), the response, $v_i.M$, of user vertex $v_i$ is a truth ($T$) with probability $v_i.P_{T|T} \in [0, 1]$;

    \item {\bf Case 2:} When $v_j.M$ is a truth ($T$), the response, $v_i.M$, of user vertex $v_i$ is a rumor ($R$) with probability $v_i.P_{R|T} \in [0, 1]$;

    \item {\bf Case 3:} When $v_j.M$ is a rumor ($R$), the response, $v_i.M$, of user vertex $v_i$ is a truth ($T$) with probability $v_i.P_{T|R} \in [0, 1]$;

    \item {\bf Case 4:} When $v_j.M$ is a rumor ($R$), the response, $v_i.M$, of user vertex $v_i$ is a rumor ($R$) with probability $v_i.P_{R|R} \in [0, 1]$;

    \item {\bf Case 5:} When there is no news from  user $v_i$'s neighbors, (initial) message $v_i.M$ posted by user $v_i$ is a truth with probability $v_i.P_{T|\emptyset} \in [0, 1]$;

    \item {\bf Case 6:} When there is no news from  user $v_i$'s neighbors, (initial) message $v_i.M$ posted by user $v_i$ is a rumor with probability $v_i.P_{R|\emptyset} \in [0, 1]$.

\end{itemize}
\noindent where $v_i.P_{T|T} + v_i.P_{R|T} = 1$, $v_i.P_{T|R} + v_i.P_{R|R} = 1$, and $v_i.P_{T|\emptyset} + v_i.P_{R|\emptyset} = 1$.
\label{def:user_behavior}
\end{definition}


\begin{table}[t]\small
\begin{center}
\caption{The Response Behavior of User $v_i$ (``$-$'' indicates no messages).}
\vspace{-2ex}
\label{tab:user_behavior}
\begin{tabular}{|c|c|c|c|}
\hline
\textbf{Case }&\textbf{Neighbor }&\textbf{User }&\textbf{Conditional} \\
\textbf{No.}&\textbf{Message, $v_j.M$}&\textbf{Response, $v_i.M$}&\textbf{Probability} \\
\hline\hline
    {\bf Case 1} & $T$ (Truth) & $T$ (Truth) & $v_i.P_{T|T}$\\\hline
    {\bf Case 2} & $T$ (Truth) & $R$ (Rumor) & $v_i.P_{R|T}$\\\hline
    {\bf Case 3} & $R$ (Rumor) & $T$ (Truth) & $v_i.P_{T|R}$\\\hline
    {\bf Case 4} & $R$ (Rumor) & $R$ (Rumor) & $v_i.P_{R|R}$\\\hline 
    \hline  
    {\bf Case 5} & $-$ & $T$ (Truth) & $v_i.P_{T|\emptyset}$\\\hline   
    {\bf Case 6} & $-$ & $R$ (Rumor) & $v_i.P_{R|\emptyset}$\\\hline 
\end{tabular}
\end{center}
\vspace{-3ex}
\end{table}
Here, $T/R$ for user responses denotes truth-/rumor-aligned behaviors derived from the response stance and historical news label.
The user response behaviors in Definition \ref{def:user_behavior} describe the reactions of user $v_i$ to truth/rumor messages from his/her neighbors. For example, Cases 1 and 4 simply copy or agree with what one's friends think; Case 2 corresponds to a potentially malicious user who posts rumors instead of conveying truths; and Case 3 is a well-behaved user who corrects rumors to truths. 

\noindent {\bf Multiple Neighbors' Influences on User Behaviors.} Definition \ref{def:user_behavior} only considers the user's reaction to the message from a single neighbor. 

Next, we would like to obtain the probability, $Pr_{WI}\{\cdot\}$, of user $v_i$'s actual response $v_i.M$ (i.e., posting $T$ or $R$ from the observation), \textit{with influences} (WI) from multiple neighbors, which is proportional to the weighted sum of conditional probabilities from neighbors $v_j$ (as shown in Table \ref{tab:user_behavior}), where weights of edges $e_{ji}$ are given by influences $e_{ji}.inf$.


\begin{eqnarray}
    &&\hspace{-4ex}Pr_{WI}\{v_i.M = R\} \propto \label{eq:eq1}\\
    &&\hspace{-4ex} \sum_{\forall v_j\in N(v_i)} \begin{cases}
    v_i.P_{R|T}\cdot e_{ji}.inf, & \text{if } v_j.M = \overline{v_i.M} = T; \\
    v_i.P_{R|R}\cdot e_{ji}.inf, & \text{if } v_j.M = v_i.M = R; \\
    0, & \text{otherwise.} \\
    \end{cases}\notag \\
    &&\hspace{-4ex} +     v_i.P_{R|\emptyset} \cdot \chi\left(\bigwedge_{\forall v_j\in N(v_i)} \left(v_j.M \ne v_i.M \bigwedge v_j.M \ne \overline{v_i.M}\right)\right);\notag 
\end{eqnarray}

\begin{eqnarray}
    &&\hspace{-4ex}Pr_{WI}\{v_i.M = T\} \propto \label{eq:eq2}\\
    &&\hspace{-4ex}\sum_{\forall v_j\in N(v_i)} \begin{cases}
    v_i.P_{T|T}\cdot e_{ji}.inf, & \text{if } v_j.M = v_i.M = T; \\
    v_i.P_{T|R}\cdot e_{ji}.inf, & \text{if } v_j.M = \overline{v_i.M}= R; \\
    0, & \text{otherwise.} \\
    \end{cases}\notag \\
    &&\hspace{-4ex}+     v_i.P_{T|\emptyset} \cdot \chi\left(\bigwedge_{\forall v_j\in N(v_i)} \left(v_j.M \ne v_i.M \bigwedge v_j.M \ne \overline{v_i.M}\right)\right),\notag
\end{eqnarray}


\noindent where $N(v_i)$ is a set of vertex $v_i$'s 1-hop neighbors, and $\chi(Z) = 1$ if variable $Z$ is true (0, otherwise).

Intuitively, Eq.~(\ref{eq:eq1}) provides the (proportional) probability that message $v_i.M$ from user $v_i$ is a rumor under the influence of his/her neighbors, which sums up the weighted conditional probabilities that one's neighbor messages $v_j.M$ agree or disagree with $v_i.M$ and the probability $v_i.P_{R|\emptyset}$ that no neighbors post relevant news, but $v_i.M$ is a rumor. The case that user $v_i$ posts a truth message $v_i.M$ is similar and is given in Eq.~(\ref{eq:eq2}). For simplicity, here, we do not normalize the weighted sum in Eqs.~(\ref{eq:eq1}) and (\ref{eq:eq2}).

\subsection{Rumor Assertion Problem Over Social Networks (RA-SN)}

\noindent {\bf Rumor/Truth Probabilities.} Given a news item $X$, assume that $S(X)$ is a set of user vertices $v_i \in V(G)$ that post/forward/comment on the news related to $X$, and $\overline{X}$ is a negation of (opposite to or disagree with) news $X$.

Below, we compute (proportional) rumor or truth probability $Pr\{X=R\}$ or $Pr\{X=T\}$, respectively, by considering the users' discriminative ability about news $X$ without the influence of one's neighbors.

\begin{eqnarray}
   &&\hspace{-2ex}Pr\{X = R\}  \\ 
   &\hspace{-2ex}\propto& \hspace{-2ex}\frac{1}{|S(X)|} \cdot \sum_{\forall v_i\in S(X)} \begin{cases}
    Pr_{WoI}\{v_i.M = R\}, & \text{if } v_i.M = X; \\
    Pr_{WoI}\{v_i.M = T\}, & \text{if } v_i.M = \overline{X}. \notag
\end{cases}
\end{eqnarray}
\begin{eqnarray}
   &&\hspace{-2ex}Pr\{X = T\}  \\ 
   &\hspace{-2ex}\propto& \hspace{-2ex}\frac{1}{|S(X)|} \cdot \sum_{\forall v_i\in S(X)} \begin{cases}
    Pr_{WoI}\{v_i.M = T\}, & \text{if } v_i.M = X; \\
    Pr_{WoI}\{v_i.M = R\}, & \text{if } v_i.M = \overline{X}. \notag
\end{cases}
\end{eqnarray}


Here, the probability, $Pr_{WoI}\{\cdot\}$, that user $v_i$'s message $v_i.M$ is truth/rumor, \textit{without influences} (WoI) from one's neighbors in $N(v_i)$, given as follows.

\begin{align}
   &Pr_{WoI}\{v_i.M = R\} \propto
   \\
   &\sum_{\forall v_j\in N(v_i)} \begin{cases}
    \frac{1}{|N(v_i)\cap S(X)|} \cdot v_i.P_{R|T}, & \text{if } v_j.M = \overline{v_i.M} = T; \\
    \frac{1}{|N(v_i)\cap S(X)|} \cdot v_i.P_{R|R}, & \text{if } v_j.M = v_i.M = R; \\
    0, & \text{otherwise.} \\
    \end{cases}\notag\\
    &+     v_i.P_{R|\emptyset} \cdot \chi\left(\bigwedge_{\forall v_j\in N(v_i)} \left(v_j.M \ne v_i.M \bigwedge v_j.M \ne \overline{v_i.M}\right)\right);\notag
\label{eq:prwoi_F}
\end{align}

\begin{align}
   &Pr_{WoI}\{v_i.M = T\}
   \propto \\ &\sum_{\forall v_j\in N(v_i)} \begin{cases}
    \frac{1}{|N(v_i)\cap S(X)|} \cdot v_i.P_{T|T}, & \text{if } v_j.M = v_i.M = T; \\
    \frac{1}{|N(v_i)\cap S(X)|} \cdot v_i.P_{T|R}, & \text{if } v_j.M = \overline{v_i.M}= R; \\
    0, & \text{otherwise.} \\
\end{cases}\notag\\
&+v_i.P_{T|\emptyset} \cdot \chi\left(\bigwedge_{\forall v_j\in N(v_i)} \left(v_j.M \ne v_i.M \bigwedge v_j.M \ne \overline{v_i.M}\right)\right).\notag
\label{eq:prwoi_T}
\end{align}


\nop{
Given a news $X$, we have the probability that $X$ is a truth/rumor below:


\begin{equation}
\begin{split}
Pr_{WoI}\{X= T\} = {} & \sum_{\forall v_i\in V(G)} \biggl( Pr_{WoI}\{v_i.M = X = T\} \\
                      & + Pr_{WoI}\{v_i.M = \overline{X} = R\} \biggr)
\end{split},
\label{eq:prwoi_X_T}
\end{equation}

\begin{equation}
\begin{split}
Pr_{WoI}\{X= R\} = {} & \sum_{\forall v_i\in V(G)} \biggl( Pr_{WoI}\{v_i.M = X = R\} \\
                      & + Pr_{WoI}\{v_i.M = \overline{X} = T\} \biggr)
\end{split}.
\label{eq:prwoi_X_R}
\end{equation}

}

We say that news $X$ is a rumor, if it holds that $Pr\{X= R\} > Pr\{X=T\}$; otherwise, news $X$ is a truth.

\vspace{0.5ex}\noindent {\bf Assertion Confidence via Entropy.} The confidence of this truth/rumor assertion can be measured by an entropy:
\begin{align}
entropy(X) = &- Pr\{X = T\} \cdot log (Pr\{X = T\}) \notag\\
&- Pr\{X = R\} \cdot log (Pr\{X = R\}).
\end{align}

A larger entropy value indicates more uncertainty of the assertion.


\vspace{0.5ex}\noindent {\bf The RA-SN Problem Statement.} Now we are ready to formalize the rumor assertion problem to decide whether or not a news item $X$ spread across social networks $G$ is a rumor.

\begin{definition} 
\textbf{(Rumor Assertion Over Social Networks, RA-SN)}
Given a social-network graph $G$, an entropy threshold $\theta \in (0, 1]$, and a query news $X$, the problem of \textit{rumor assertion over social networks} (RA-SN) asserts whether or not news $X$ is a rumor $R$. 

News $X$ is asserted as a rumor $R$, if it holds that:
\begin{itemize}
     \item {\bf(Rumor/Truth Probability)} $Pr\{X= R\} > Pr\{X=T\}$, and;
     \item {\bf(Assertion Confidence)} $entropy(X) \leq \theta$.
\end{itemize}


\label{def:TRA-SN}
\end{definition}




Table~\ref{tab:notations} depicts the commonly used notations and their descriptions in this paper.

\begin{table}[t]\small
\begin{center}
\caption{Symbols and Descriptions}
\label{tab:notations}
\vspace{-2ex}
\begin{tabular}{|p{1.8cm}|p{6.3cm}|}
\hline
\textbf{Symbol}&\textbf{Description} \\
\hline\hline
    $G$ & a temporal social-network graph\\\hline
    $V(G)$ & a set of vertices $v_i$ \\\hline
    $E(G)$ & a set of edges $e_{ij}$ \\\hline
    $v_i$ & a vertex in graph $G$\\\hline
    $e_{ij}$ (or $e_{v_i,v_j}$) & an edge in graph $G$\\\hline
    $e_{ij}.inf$ & the influence of vertex $v_i$ on vertex $v_j$ via edge $e_{ij}$\\\hline
    $\phi(G)$ &  a mapping function $V(G)\times V(G)\rightarrow E(G)$ \\\hline
    $v_i.M$ & a message posted/forwarded/commented by user $v_i$ \\\hline
    $\mathbf{P}$ & the 6-dimensional probability matrix \\\hline
    $\Delta$ & the correction of the probability matrix $\mathbf{P}$ \\\hline
\end{tabular}
\end{center}
\end{table}

\begin{algorithm}[!t]
    \caption{\bf The RA-SN Processing Framework}
    \label{alg:ra-sn}
    \setcounter{AlgoLine}{0}
    \KwIn{
        \romannumeral1) news set $\mathcal{S}$ and comments set $\mathcal{C}$,
        \romannumeral2) social-network graph $G$,
        \romannumeral2) a set, $Q$, of assertion news $X$, and 
        \romannumeral3) an entropy threshold $\theta$
    }
    \KwOut{
        assertion results $\hat{Y}_X$
    }
    
    \tcp{data pre-computation phase}

    pre-compute an initial probability matrix $\mathbf{P}$, an evidence quantity matrix $\mathbf{A}$, and a posterior variance matrix $\mathbf{V}$

    \tcp{embedding learning and index construction phase} 
    
    train a GNN encoder, Enc($\cdot$), and decoder, Dec($\cdot$)

    generate reconstructed behavioral probability representations $\mathbf{P}_{rec}$
    
    construct an embedding tree index $\mathcal{I}$ over the Hellinger embeddings $\Phi(\mathbf{P}_{rec})$

    \tcp{online rumor assertion phase}
    
    \For{each query assertion news  $X \in Q$}{

        extract the users involved from the cascade of $X$


        refine insufficient evidence probability groups using top-$k$ similar historical users retrieved from $\mathcal{I}$

        aggregate evidence data and obtain assertion result $\hat{Y}_X$ of $X$ w.r.t. rumor probability and entropy threshold $\theta$
        
        \Return $\hat{Y}_X$
    }

\end{algorithm}

\nop{

{\color{red}
\noindent * learn $v_i.P_{T|T}$, $v_i.P_{R|T}$, $v_i.P_{T|R}$, and $v_i.P_{R|R}$ in Table \ref{tab:user_behavior} (real-world data can be noisy, inaccurate, and incomplete)\\

\noindent * learn/estimate the edge influence $e_{ij}.inf$ or using IC/LT model\\

\noindent * decide a news $X$ or message $v_i.M$ falls into some category $c_x$\\

\noindent * compare news $X$ with message $v_i.M$ (post, forwarded message, or reply/comments), and evaluate whether they are related, whether they have the same or opposite contents/opinions
}

}

\section{The RA-SN Processing Framework}
\begin{figure*}[!t]
    \centering
    \includegraphics[width=0.95\linewidth]{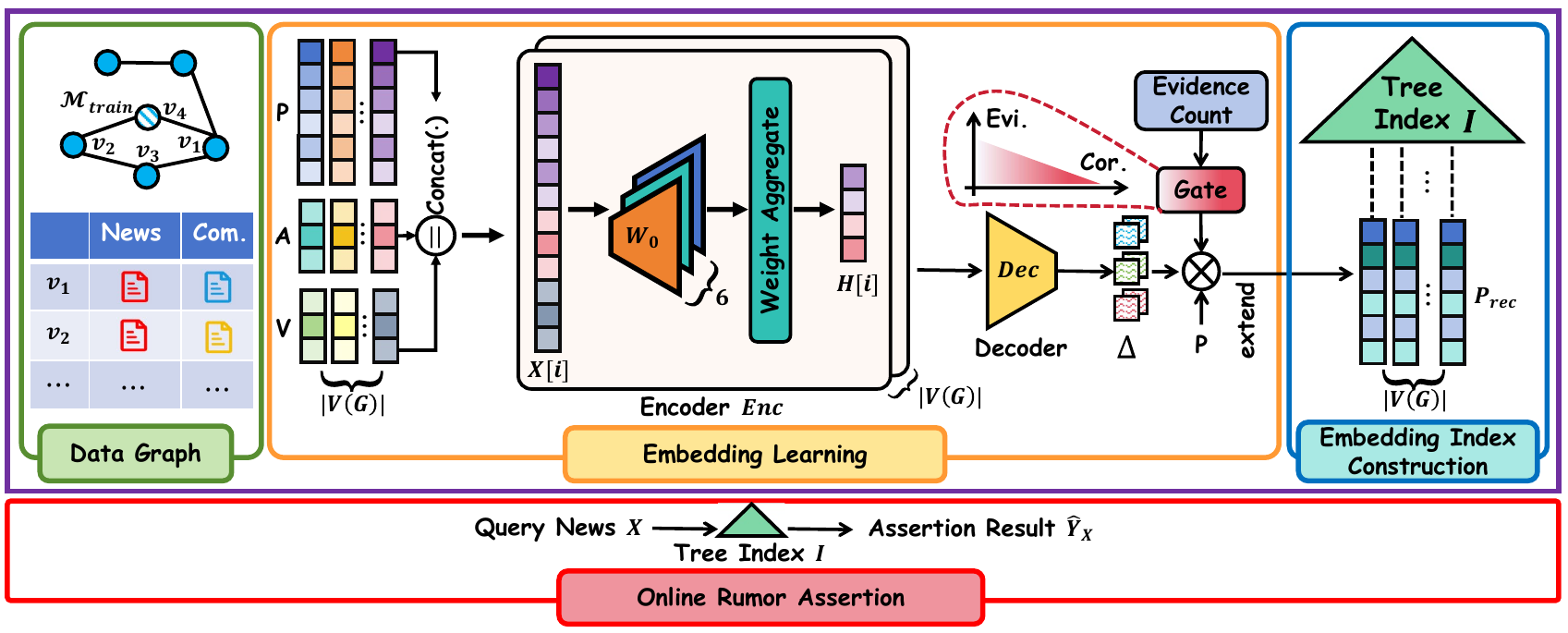}
    \caption{The framework and workflow for the rumor assertion over social networks (RA-SN).}
    \label{fig:workflow1}
\end{figure*}

Algorithm~\ref{alg:ra-sn} illustrates a framework for processing the RA-SN problem, which consists of three phases: \textit{data pre-computation}, \textit{embedding learning and index construction}, and \textit{online rumor assertion phases}. Figure~\ref{fig:workflow1} provides a visual workflow of the pseudo code in Algorithm~\ref{alg:ra-sn}.

As shown in Algorithm~\ref{alg:ra-sn} and Figure~\ref{fig:workflow1}, we first preprocess the social-network graph $G$ to obtain an initial probability matrix $\mathbf{P}$ (including a 6-dimensional probability vector per user), an evidence quantity matrix $\mathbf{A}$ (including a 3-dimensional evidence quantity vector per user), and a posterior variance matrix $\mathbf{V}$ (including a 3-dimensional posterior variance vector per user, line 1). 


Then, in the embedding learning and index construction phase, we train a GNN-based encoder, Enc($\cdot$), and a decoder, Dec($\cdot$), to generate the reconstructed behavioral probability representations $\mathbf{P}_{rec}$ for graph $G$, and build an embedding tree index $\mathcal{I}$ over the Hellinger embeddings $\Phi(\mathbf{P}_{rec})$ for the rumor assertion task (lines 2-4).

In the online rumor assertion phase, for each query assertion news $X \in Q$, we first extract the users involved from the cascade of $X$ (lines 5-6). Next, we identify the probability groups with insufficient historical evidence and retrieve the top-$k$ most similar historical users from the embedding index $\mathcal{I}$ to refine these groups for each participating user (line 7).
Finally, we aggregate the evidence to determine and return the assertion result $\hat{Y}_X$ for $X$, based on the rumor probability and entropy threshold $\theta$ (lines 8-9).


\begin{figure}[!t]
    \centering
    \includegraphics[width=0.75\linewidth]{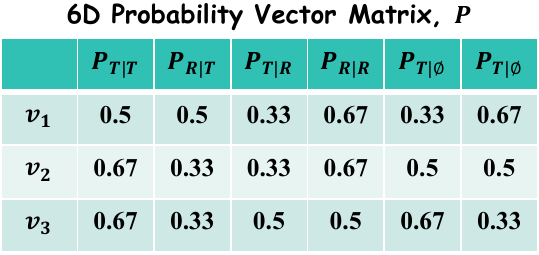}
    \caption{Example of 6-dimensional Probability Matrix $P$.}
    \label{fig:P}
\end{figure}

\subsection{The Pre-Computation Phase}

For each comment $c$ in the training set $\mathcal{D}_{train}$, we first infer its stance $s\in \{S, O\}$ toward the corresponding news claim together with a confidence score, where $S$ indicates agreement with the statement in $c$, and $O$ represents disagreement or questioning of the statement in $c$.
We employ an off-the-shelf nli-deberta-v3-small~\cite{he2021debertav3} model to infer the stance of each comment toward its corresponding news claim, together with the associated prediction confidence, without any task-specific fine-tuning on our datasets.
Given the ground-truth label $y\in\{T,R\}$, we determine whether the user's judgment is correct by
\begin{equation}
C_u=\mathbb{I}\big[(s_u=S\land y=T)\vee(s_u=O\land y=R)\big].
\end{equation}

Next, we aggregate confidence-weighted historical interaction evidence. For an interaction event $e=(u,v)$, the joint stance confidence is defined as
$q_e=\sqrt{q_u^e q_v^e}$.
According to the correctness states of users $u$ and $v$, each interaction contributes to one of four evidence types, i.e., $\mathcal{N}_{T|T}$, $\mathcal{N}_{R|T}$, $\mathcal{N}_{T|R}$, and $\mathcal{N}_{R|R}$. In addition, we count the number of real and rumor news items posted by each user, denoted by $\mathcal{N}_{T|\emptyset}$ and $\mathcal{N}_{R|\emptyset}$.

Based on these statistics, we organize the six evidence types into three complementary groups,
$(\mathcal{N}_{T|T},\mathcal{N}_{R|T}),
(\mathcal{N}_{T|R},\mathcal{N}_{R|R}),
(\mathcal{N}_{T|\emptyset},\mathcal{N}_{R|\emptyset})$,
and apply Beta smoothing to estimate the corresponding probabilities. For the $k$-th group with positive and negative evidence $\mathcal{N}_k^+$ and $\mathcal{N}_k^-$, respectively,
\begin{equation}
P_k=\frac{\mathcal{N}_k^+ +1}
{\mathcal{N}_k^+ +\mathcal{N}_k^-+2}.
\end{equation}
Accordingly, the 6-dimensional probability representation of user $u$ is
\begin{equation}
\mathbf{P}[u]=
[P_{T|T},P_{R|T},P_{T|R},P_{R|R},
P_{T|\emptyset},P_{R|\emptyset}].
\end{equation}

Since similar probability values may arise from substantially different amounts of historical evidence, we further encode evidence sufficiency and uncertainty. For each evidence group $k$, we define
\begin{equation}
A_k=\log(1+\mathcal{N}_k^++\mathcal{N}_k^-),
\label{eq:three-evidence}
\end{equation}
and measure posterior uncertainty using the variance of the corresponding Beta distribution:
\begin{equation}
V_k=
\frac{\alpha_k\beta_k}
{(\alpha_k+\beta_k)^2(\alpha_k+\beta_k+1)},
\end{equation}
where $\alpha_k=\mathcal{N}_k^++1$ and $\beta_k=\mathcal{N}_k^-+1$.

Finally, we obtain the 3-dimensional evidence quantity vector $\mathbf{A}$ and uncertainty vector $\mathbf{V}$, and concatenate them with $\mathbf{P}$ to construct the 12-dimensional initial user feature:
\begin{equation}
\mathbf{F}[u]=[\mathbf{P}[u],\mathbf{A}[u],\mathbf{V}[u]].
\end{equation}
We further generate a training mask $\mathcal{M}_{train}$ based on users' accumulated valid historical evidence to exclude extremely sparse users from reconstruction supervision.

\subsection{Evidence-Aware Probability Reconstructor and Training Details}

\begin{algorithm}[!t]
    \caption{\bf Evidence-aware Probability Reconstructor}
    \label{alg:ed}
    \setcounter{AlgoLine}{0}
    \KwIn{
        \romannumeral1) a user feature matrix $\mathbf{F}$,
        \romannumeral2) an evidence count matrix $\mathcal{N}$,
        \romannumeral3) a data graph $G$,
        \romannumeral4) a training mask $\mathcal{M}_{train}$,
        \romannumeral5) a mask rate $\rho$, and
        \romannumeral6) a maximum number of training epochs $Epochs$
    }
    \KwOut{
        the reconstructed user probability matrix $\mathbf{P}_{rec}$
    }
    \tcp{model initialization}

    initialize GNN-based encoder $\text{Enc}(\cdot)$ and probability correction decoder $\text{Dec}(\cdot)$

    filter training nodes $V_{train} \leftarrow \{u \in V(G) \mid \mathcal{M}_{train}[u] = \text{True}\}$

    \tcp{reconstructor training}

    \For{\text{epoch} =1 \KwTo $Epochs$}{

        \For{each sampled batch $\mathcal{B}$ centered on $V_{train}$}{

            randomly select masked nodes $\mathcal{B}_
            {mask}$ from the central training nodes with rate $\rho$

            construct masked features
            $\widetilde{\mathbf{F}}_{\mathcal{B}}$ and
            $\widetilde{\mathcal{N}}_{\mathcal{B}}$

            $\mathbf{H}
            \leftarrow
            \text{Enc}(\widetilde{\mathbf{F}}_{\mathcal{B}},
            G_{\mathcal{B}})$

            $\boldsymbol{\Delta}\leftarrow\text{Dec}(\mathbf{H})$
            
            \tcp{evidence correction}

            compute evidence gates $\mathbf{g}$ from  $\widetilde{\mathcal{N}}_{\mathcal{B}}$

            extract base probabilities $\mathbf{P}_{base}$ from
            $\widetilde{\mathbf{F}}_{\mathcal{B}}$
            

             $\mathbf{P}_{rec}^{\mathcal{B}}
            \leftarrow
            \text{Extend}\!\left(
            \sigma(
            \operatorname{logit}(\mathbf{P}_{base})
            +\mathbf{g}\boldsymbol{\Delta})
            \right)$

            compute binomial reconstruction loss on
            $\mathcal{B}_{mask}$ using original evidence counts

            update model parameters by backpropagation
            
        }
        
    }

    \For{each sampled batch $\mathcal{B}$ over all users}{

        $\mathbf{P}_{rec}^\mathcal{B} \leftarrow \text{ModelForward}(\mathbf{F}_{\mathcal{B}}, \mathcal{N}_{\mathcal{B}}, G_{\mathcal{B}})$
    
    }
    
    \Return $\mathbf{P}_{rec}$
\end{algorithm}

We design an evidence-aware probability reconstructor (EPR) that enhances and generalizes the representation of individual behavior probabilities through the structural semantics between users.
Unlike methods that directly use news ground-truth labels to supervise user representation learning,
EPR is trained through a masked evidence reconstruction task.
Specifically, EPR randomly hides certain pieces of evidence to fully understand user features and uses their neighbor information to recover the behavioral distribution, thereby learning the potential correlation between user historical evidence and structural semantics. Algorithm~\ref{alg:ed} shows the training details of EPR. 

\noindent\textbf{Graph Encoding and Decoding}. First, we filter users with sufficient historical evidence from the full user set based on the training mask $\mathcal{M}_{train}$: $\{u \in V(G) \mid \mathcal{M}_{train}[u] = \text{True}\}$ (line 2 of Algorithm~\ref{alg:ed}).
In order to train on a large-scale graph, in each training epoch, we use the user $u \in V_{train}$ as the central node and construct a local subgraph $\mathcal{B}$ through neighbor sampling.
For each batch, we randomly select users at a proportion of $\rho$ from the central training nodes to form a mask set $\mathcal{B}_{mask}$ (lines 3-5 of Algorithm~\ref{alg:ed}). 
For those users, we replace the 6-dimensional probabilistic features with 0.5, set the 3-dimensional evidence amount to 0, and set the 3-dimensional posterior variance to 1/12, corresponding to the uniform prior. We set the corresponding 6-dimensional evidence counts to 0. Thus, we obtain the masked user features $\widetilde{\mathbf{F}}_{\mathcal{B}}$ and evidence counts $\widetilde{\mathcal{N}}_{\mathcal{B}}$ (line 6 of Algorithm~\ref{alg:ed}).

We employ two behavior convolution layers (BCL) as the encoder $\text{Enc}(\cdot)$ for EPR. The encoder $\text{Enc}$ explicitly models different probability types and their influence weights, rather than using a uniform message transformation for all user interactions.
For a connected edge from user $j$ to user $i$, let its probability type be $r_{ji}$ and its influence weight be $w_{ji}$, then the neighbor message can be represented as:
$w_{ji}\mathbf h_j^{(l)}\mathbf W_{r_{ji}}^{(l)}$,
where $\mathbf W_{r_{ji}}^{(l)}$ is the independent parameter matrix corresponding to the probability type, and $\mathbf h_j^{(l)}$ is the representation of node $j$ in the $l$-th layer.

For node $i$, all incoming edge messages are normalized and aggregated according to the edge weights, and together with the node's own representation, they generate the next layer representation:
\begin{eqnarray}
    \mathbf h_i^{(l+1)} = \sigma(\mathbf W_{self}^{(l)}\mathbf h_i^{(l)} +\frac{\sum_{j\in\mathcal N(i)}w_{ji}\mathbf h_j^{(l)}\mathbf W_{r_{ji}}^{(l)}}{\max(1,\sum_{j\in\mathcal N(i)}w_{ji})}),
\end{eqnarray}
where $\sigma$ is the ReLU activation function. Additionally, we employ residual connections between graph layers to mitigate over-smoothing.

After two layers of information propagation, we can obtain the user hidden representation matrix $\mathbf{H}$ (line 7 of Algorithm~\ref{alg:ed}). Then, we design a probabilistic correction decoder $\text{Dec}(\cdot)$ to map it into three structural corrections $\boldsymbol{\Delta}\in \mathbb{R}^{3}$ (line 8 of Algorithm~\ref{alg:ed}): 
\begin{eqnarray}
    \text{Dec}(\mathbf{H})\rightarrow
\boldsymbol{\Delta}\in\mathbb R^{|\mathcal B|\times3}.
\end{eqnarray}
The three correction dimensions correspond to three complementary evidence groups (i.e., Eq.~\ref{eq:three-evidence}).

\noindent\textbf{Evidence Probability Correction}.
Since the two probabilities in each group are complementary (e.g., $P_{T|T}[u]+P_{R|T}[u]=1$), the EPR model only needs to directly correct the first probability in each group (i.e., $P_{T|T}, P_{T|R}$, and $P_{T|\emptyset}$).

To control how much structural information modifies the original probability, we further construct an evidence gate based on the amount of historical evidence from users. For the $k$-th evidence group, the gating coefficient is:
\begin{eqnarray}
    g_k = 1/\sqrt{(1+\mathcal{N}_k^++\mathcal{N}_k^-)}.
\end{eqnarray}
Intuitively, the gating mechanism adaptively adjusts the strength of the structural correction based on the sufficiency of historical evidence. In other words, the more sufficient the evidence, the weaker the structural correction; conversely, the weaker the evidence, the stronger the correction.

Then, the correction probability of the $k$-th evidence group is:
\begin{eqnarray}
    \hat{P}_k = \sigma(logit(P_k)+\mathbf{g}_k\boldsymbol{\Delta}_k).
\end{eqnarray}
Finally, we use an extend operation to extend the three corrected probabilities into a 6-dimensional reconstructed user behavior representation:
\begin{eqnarray}
    \mathbf{P}_{rec} = [\hat{P}_1, 1\text{-}\hat{P}_1,\hat{P}_2, 1\text{-}\hat{P}_2,\hat{P}_3, 1\text{-}\hat{P}_3].
\end{eqnarray}
Compared with directly using the output of a 6-dimensional MLP and applying grouped softmax, the EPR model explicitly preserves the probability estimates from the data preprocessing stage and uses the behavior network only to learn structural corrections, thereby avoiding over-smoothing of reliable historical evidence (lines 9-13 of Algorithm~\ref{alg:ed}).

\noindent\textbf{Reconstructor Loss Function}. 
To enable the model to recover users' potential behavioral distribution from structural semantics while making the most of the sufficient observation signals provided by the original evidence count, we design a binomial negative log-likelihood based on the historical evidence count as the reconstruction loss. 

Specifically, for the $k$-th evidence group of user $u$, the corresponding reconstruction loss $\mathcal{L}$ is:
\begin{eqnarray}
    \mathcal{L}_{u,k} = -[ N_{k}^{+}[u]\log \hat P_{k}[u]+N_{k}^{-}[u]\log(1-\hat P_{k}[u])].
\end{eqnarray}

During training, the reconstruction loss $\mathcal{L}_{rec}$ is calculated only for the set of randomly masked center nodes $\mathcal B_{mask}$ in the current batch. 
We sum the losses of the three evidence groups and normalize them according to the total amount of historical evidence for these nodes:
\begin{eqnarray}
   \mathcal{L}_{rec} =  \frac{
\displaystyle
\sum_{u\in\mathcal B_{mask}}
\sum_{k=1}^{3}\mathcal{L}_{u,k}
}{
\displaystyle
\sum_{u\in\mathcal B_{mask}}
\sum_{k=1}^{3}
\left(
N_{k}^{+}[u]+N_{k}^{-}[u]
\right)
}.
\end{eqnarray}

Finally, we jointly optimize the graph encoder and the probability correction decoder by minimizing $\mathcal{L}_{rec}$.
After model training, we perform forward propagation on all users using the complete user features and graph to obtain the final 6-dimensional reconstructed user probability matrix $\mathbf{P}_{rec}$ (lines 14-16 of Algorithm~\ref{alg:ed}).

\noindent\textbf{Embedding Tree Indexing Mechanism}.
In this subsection, we detail how to construct a tree embedding index $\mathcal{I}$ on a data graph based on user behavior representations precomputed by the EPR model to support online RA-SN processing. 
To measure the embedding representation, we introduce the Hellinger mapping for the EPR 6-dimensional representation defined as $\Phi(\mathbf{P}_u)$:
\begin{eqnarray}
\Phi(\mathbf{P}_u)
=
\frac{1}{\sqrt{2}}
\left[
\sqrt{P_{T|g}},\sqrt{P_{R|g}}
\right]_{g\in\{T,R,\emptyset\}}.
\end{eqnarray}
The $\Phi(\mathbf{P}_u)$ transforms the Hellinger distance~\cite{cieslak2012hellinger} between probability distributions into Euclidean distance~\cite{gower1985properties} in the embedding space, thereby enabling non-leaf nodes to characterize their child users via center and covering radius.

We construct a hierarchical index $\mathcal{I}$. The nodes of $\mathcal{I}$ are divided into two classes: leaf nodes and non-leaf nodes.

\underline{\textit Leaf Nodes}: each leaf node $\mathcal{N}$ contains a historical user set $U_{\mathcal{N}}\subseteq V(G)$. For each user $u \in U_{\mathcal{N}}$, the leaf node stores the following precomputed data:
\begin{itemize}
    \item user ID, $u$;
    \item EPR 6-dimensional representation, $\mathbf{P}_u$;
    \item the evidence counts of user $u$, $\mathbf{N}_u$.
\end{itemize}

\underline{\textbf Non-leaf Nodes}:  each non-leaf node $\mathcal{N}$ represents a coverage area for all child users and contains the following aggregated data:
\begin{itemize}
    \item a pointer to child nodes, $Ptr_\mathcal{N}$;
    \item an embedding center, $\mathbf{c_{\mathcal{N}}}=\frac{1}{|U_{\mathcal{N}}|}\sum_{u\in U_\mathcal{N}}\Phi(\mathbf{P}_u)$;
    \item a covering radius, $r_{\mathcal{N}}$;
    \item the lower and upper vector $\mathbf{l}_{\mathcal{N}}=\min_{u\in U_{\mathcal{N}}}\Phi(\mathbf{P}_u)$ and $\mathbf{U}_{\mathcal{N}}=\max_{u\in U_{\mathcal{N}}}\Phi(\mathbf{P}_u)$ of axis-aligned bounding box;
    \item the number of users in the current nodes $n_{\mathcal{N}}$.
\end{itemize}

\underline{\textbf{Index Construction Processing}}:
We construct the index $\mathcal{I}$ using a top-down approach. 
The root node covers all users in the data graph $G$ and is inserted into the processing queue $Q$.
For each node $\mathcal{N}$ removed from the queue, we compute its center, covering radius, and axis-aligned bounding box.
If the current node size does not exceed $L$, we mark it as a leaf node.
Otherwise, we select the mapped dimension $j$ with the largest variance and sort the users along that dimension. 
Then, we partition the sorted user sequence into $fanout_{\mathcal{N}}$ contiguous segments of approximately equal size and create a child node for each segment. 
Until all nodes satisfy the fanout size and leaf-capacity constraint, we return the index $\mathcal{I}$.

\noindent\textbf{Joint Distance Lower Bound Pruning}. 
During top-$K$ retrieval, we maintain the current $K$ nearest users in a max heap and let $\tau$ denote the distance of the current $K$-th nearest user. To avoid traversing index nodes that cannot contain a better candidate, we derive a distance lower bound between the query embedding $\mathbf q$ and all user embeddings covered by an index node $\mathcal N$.

\begin{lemma}
    \label{lemma:bound_pruning} 
    \textbf{(Joint Distance Lower Bound Pruning)} 
    Given a query representation $\mathbf{q}$ and an index node $\mathcal{N}$, we define the corresponding joint distance lower bound as $LB_{joint}(\mathcal{N},\mathbf{q})=\max\left\{LB_{ball}(\mathcal{N},\mathbf{q}), LB_{box}(\mathcal{N},\mathbf{q})\right\}$, where $LB_{ball}(\mathcal{N},\mathbf{q})=\max(0,\|\mathbf{q}-\mathbf{c}_{\mathcal{N}}\|_2-r_{\mathcal{N}}), LB_{box}(\mathcal{N},\mathbf{q}) = \sqrt{\sum_j\left[\max\left\{\mathbf{l}_{\mathcal{N}}[j]-\mathbf{q}[j],0,\mathbf{q}[j]-\mathbf{h}_{\mathcal{N}}[j]\right\}\right]^2}$.
    If$LB_{joint}(\mathcal{N},\mathbf{q})>\tau$, then node $\mathcal{N}$ and all its descendants can be safely pruned.
\end{lemma}

\begin{proof}
    For any user $u\in U_{\mathcal{N}}$, the triangle inequality in the Hellinger representation space gives
    \begin{eqnarray}
    \|\mathbf{q}-\Phi(\mathbf{P}_u)\|_2
    \geq
    \|\mathbf{q}-\mathbf{c}_{\mathcal{N}}\|_2
    -
    \|\Phi(\mathbf{P}_u)-\mathbf{c}_{\mathcal{N}}\|_2.
    \end{eqnarray}
    By the definition of the covering radius,
    $\|\Phi(\mathbf{P}_u)-\mathbf{c}_{\mathcal{N}}\|_2\leq r_{\mathcal{N}}$; therefore,
    \begin{eqnarray}
    \|\mathbf{q}-\Phi(\mathbf{P}_u)\|_2
    \geq LB_{ball}(\mathcal{N},\mathbf{q}).
    \end{eqnarray}
    
    Moreover, for every dimension $j$,
    $\Phi(\mathbf{P}_u)[j]\in
    [\mathbf{l}_{\mathcal{N}}[j],\mathbf{h}_{\mathcal{N}}[j]]$.
    Thus, the coordinate-wise distance between the query and the user representation is at least $\delta_j(\mathcal{N},\mathbf{q})$, which gives
    \begin{eqnarray}
    \|\mathbf{q}-\Phi(\mathbf{P}_u)\|_2
    \geq
    LB_{box}(\mathcal{N},\mathbf{q}).
    \end{eqnarray}
    Since both $LB_{ball}$ and $LB_{box}$ are valid lower bounds, their maximum remains a valid lower bound, i.e.,
    \begin{eqnarray}
    \|\mathbf{q}-\Phi(\mathbf{P}_u)\|_2
    \geq
    LB_{joint}(\mathcal{N},\mathbf{q}).
    \end{eqnarray}
    If $LB_{joint}(\mathcal{N},\mathbf{q})>\tau$, then
    $\|\mathbf{q}-\Phi(\mathbf{P}_u)\|_2>\tau$ for every
    $u\in U_{\mathcal{N}}$. Consequently, no user covered by $\mathcal{N}$ can improve the current top-$k$ result, and the node together with all its descendants can be safely pruned.
\end{proof}

\subsection{Online RA-SN Processing}
In this section, we detail the RA-SN online processing. Algorithm~\ref{alg:RA-SN-framework} summarizes the online processing.

For a query news item $X$ and its observed cascade, RA-SN first extracts the participating users and messages (line 1).
For users with insufficient historical evidence, we retrieve behaviorally similar historical users from the embedding tree $\mathcal{I}$ to complete their EPR representations (lines 2-9).
We then aggregate the completed user evidence to estimate the truth and rumor probabilities of $X$ and produce the final assertion result (lines 10-16).

For a participating user $v_i$, let $A_{i,g}$ denote the historical evidence amount of condition group $g\in\{T,R,\emptyset\}$. If $A_{i,g}<\delta$, intuitively, the corresponding EPR probability is considered to lack sufficient support and has insufficient features.
We therefore query the embedding tree $\mathcal{I}$ using $\mathbf P_i$ and retrieve the top-$k$ behaviorally similar historical users $\mathcal{K}_i$ according to Lemma~\ref{lemma:bound_pruning}. 

For each retrieved user $v_j\in\mathcal K_i$, its contribution under group $g$ is weighted by behavioral similarity and historical evidence:
\begin{equation}
    w_{ij,g}=\exp\left(-d_H(\mathbf P_i,\mathbf P_j)^2\right)\sqrt{\frac{\exp(A_{j,g})}{\exp(A_{j,g})+2}},
\end{equation}
where $d_H(\mathbf P_i,\mathbf P_j)$ is the Hellinger distance between their behavioral representations.
Thus, behaviorally closer users with stronger evidence contribute more.

The retrieved users are then aggregated to estimate the behavioral probabilities of $v_i$ under group $g$:
\begin{equation}
    \bar P_{T|g}^{i}=
    \frac{\sum_{v_j\in\mathcal K_i} w_{ij,g}P_{T|g}^{j}}{\sum_{v_j\in\mathcal K_i}w_{ij,g}},
    \bar P_{R|g}^{i}=1-\bar P_{T|g}^{i}.
\end{equation}

For an insufficient evidence group $g$, i.e., $A_{i,g}<\delta$, its EPR probability is completed as
\begin{eqnarray}
\widetilde P_{T|g}^{i}=
\lambda_{i,g}P_{T|g}^{i}
+
(1-\lambda_{i,g})\bar P_{T|g}^{i},
\widetilde P_{R|g}^{i}=
1-\widetilde P_{T|g}^{i},
\label{eq:PPP}
\end{eqnarray}
where $\lambda_{i,g}=
\frac{A_{i,g}}{A_{i,g}+\eta}$ denotes the proportion of original information retained, and $\eta$ controls the contribution of the retrieved historical users during representation completion.
Hence, less historical evidence leads to a larger contribution from similar historical users.

If $A_{i,g}\geq\delta$, no completion is performed:
\begin{equation}
\widetilde P_{T|g}^{i}=P_{T|g}^{i},
\widetilde P_{R|g}^{i}=P_{R|g}^{i}.
\label{eq}
\end{equation}

After completing all insufficient evidence groups, we obtain $\widetilde{\mathbf P}_i$ for each participating user.
RA-SN combines these completed behavioral representations with the observed message stances, aggregates the resulting user evidence, and obtains $\Pr{X=T}$ and $\Pr{X=R}$ for the final rumor assertion.

\begin{algorithm}[!t]
    \caption{\bf Online RA-SN Rumor Assertion}
    \label{alg:RA-SN-framework}
    \setcounter{AlgoLine}{0}
    \KwIn{
        \romannumeral1) EPR representations $\mathbf{P}_{rec}$,
        \romannumeral2) evidence counts $\mathcal{N}$,
        \romannumeral3) embedding Tree $\mathcal{I}$,
        \romannumeral4) query news $X$, and its cascade information $C_X$, and
        \romannumeral5) entropy threshold $\theta$
    }
    \KwOut{
        assertion result $\hat{Y}_X$ and entropy $H(X)$
    }

    \tcp{participant extraction}

    extract participating users, messages, stances, and propagation edges from $C_X$


    \For{each historically visible participant $v_i\in S(X)$}{

        $\widetilde{\mathbf P}_i\leftarrow\mathbf P_i$

        \If{$\exists g\in\{T,R,\emptyset\}$ such that $A_{i,g}<\delta$}{
        
            query $\mathcal{I}$ once for the exact top-$k$ similar historical   users $\mathcal{K}_i$ with Lemma~\ref{lemma:bound_pruning}
        }

        \For{each $g \in \{T,R,\emptyset\}$ with $A_{i,g}<\delta$}{

            compute $\bar P_{T|g}^{i}$ 

            update $(\widetilde P_{T|g}^{i},\widetilde P_{R|g}^{i})$
        
        }

        construct the valid event set $\mathcal E_i(X)$ from $C_X$

        
            
            

    }


    normalize $Pr\{X=T\}$ and $Pr\{X=R\}$

    compute $entropy(X)$


    

    
    \If{$entropy(X)>\theta$}{
        $\hat{Y}_X \leftarrow \textsc{Uncertain}$
    }
    \Else{
        $\hat{Y}_X \leftarrow R$ if $Pr\{X\text{=}R\}>Pr\{X\text{=}T\}$, otherwise $\hat{Y}_X \leftarrow T$
    }

    \Return $(\widehat Y_X,entropy(X))$
    
\end{algorithm}

\section{Experimental Evaluation}

\subsection{Experimental Settings}
\noindent\textbf{Graph Data Sets.} We use two publicly available real-world rumor assertion data sets, Fakeddit~\cite{nakamura2020fakeddit} and PHEME~\cite{kochkina2018pheme},
and three synthetic graph data sets to evaluate the effectiveness of our RA-SN processing approach. The details of real-world data sets are depicted in Table~\ref{tab:data sets}. 
For synthetic graph data, we generate three types of synthetic graphs, RA-Uni, RA-Gau, and RA-Zipf. 
RA-Uni, RA-Gau, and RA-Zipf sample 6-dimensional user behavior probabilities from Uniform, Gaussian, and Gaussian distributions, respectively. For user out-degree and cascade size, RA-Uni and RA-Gau follow a Zipf distribution with $\alpha\text{=}1.5$, while RA-Zipf adopts a heavier-tailed Zipf distribution with $\alpha\text{=}1.2$.
Table~\ref{tab:parameters} shows our parameter settings, where default parameters are highlighted in bold.

\noindent\textbf{Baselines.} In our experiments, we compare our RA-SN approach with six baselines, including BiGCN~\cite{bian2020rumor}, UPFD~\cite{dou2021user}, GLAN~\cite{yuan2019jointly}, DUCK~\cite{tian2022duck}, GAT~\cite{velivckovic2017graph}, and RAW (i.e., an RA-SN variant that uses user probability features solely derived from raw historical statistics without representation reconstruction and completion). 
To ensure a fair comparison with the binary-classification baselines, we set $\theta = 1$ in our main analysis. This adjustment allows RA-SN to produce a truth or rumor prediction for every test news item, resulting in 100\% prediction coverage. Additionally, we vary $\theta$ to assess the trade-off between the MF1 score and coverage for the confidence-aware selective assertion.


\noindent\textbf{Evaluation Metrics.} We evaluate the rumor assertion task in terms of \textbf{Macro F1} (MF1), \textbf{F1-Truth}, \textbf{F1-Rumor}, and \textbf{Accuracy} (ACC) scores. Higher values indicate better performance. Moreover, we also report the time/space costs of our RA-SN processing framework.


\noindent\textbf{Deafult Parameters.} We stratify the news by category into an (80\%) training set, a (10\%) validation set, and a (10\%) test set. For EPR, we employ a 32-dimensional hidden layer, a dropout rate of 0.1, and 30\% random evidence masking.
We train EPR for 30 epochs with a batch size of 1,024, sampling 10 neighbors per layer. We use the Adam optimizer with a learning rate of $10^{-3}$ and a weight decay of $5 \times 10^{-4}$.
We set the fanout size and leaf capacity of the embedding tree to 4 and 256, respectively. We average all experiments over 5 model seeds and report the sample standard deviation. All historical statistics, $G$, EPR representations, and index $\mathcal I$ are constructed using only the training split.

All experiments were conducted on a PC equipped with an NVIDIA RTX PRO 6000 GPU with 96 GB of memory and an Intel Core i9-14900K with 192GB of RAM.

\begin{table}[t]
\begin{center}
\caption{Statistics of the Tested Real-World Rumor Assertion Data Sets.}
\label{tab:data sets}
\resizebox{\linewidth}{!}{
\begin{tabular}{|l||l||l||l||l||l|}
\hline
\textbf{Name} & \textbf{\# of Truth} & \textbf{\# of Rumor} & \textbf{\# of Comm.} & $\bm{|V(G)|}$ & $\bm{|E(G)|}$ \\
\hline\hline
    $Fakeddit$ & 268,908 & 413,753 & 8,912,601 & 1,452,068 & 11,884,048 \\\hline
    $PHEME$ & 4,022 & 2,402 & 97,909 & 41,294 & 113,098 \\\hline
\end{tabular}
}
\end{center}
\vspace{-3ex}
\end{table}

\begin{table}[t]
\begin{center}
\caption{Parameter Settings.}
\label{tab:parameters}
\resizebox{\linewidth}{!}{
\begin{tabular}{|p{5cm}||l|}
\hline
\textbf{Parameters}&\textbf{Values} \\
\hline\hline
    the entropy threshold, $\theta$ & 0.4, {\bf0.5}, 0.6, 0.7 \\\hline
    \# of GNN sampled neighbors, $K_g$ & 5, {\bf 10}, 15, 20 \\\hline
    the training masking ratio, $\rho$ & 0.1, {\bf 0.3}, 0.5, 0.8 \\\hline
    the embedding dimension, $d$ & 16, {\bf 32}, 128, 256 \\\hline
    \# of top retrieved nodes, $K_r$ & 5, {\bf 10}, 20, 30 \\\hline
    \# of synthetic graph nodes, $|V(G)|$ &  {\bf 5K}, 25K, 50K, 100K, 1M 
    \\\hline
\end{tabular}
}
\end{center}
\vspace{-3ex}
\end{table}

\begin{table*}[t]
    \centering
    \small
    \begin{tabular}{c c | c c c c c c c}
    \toprule
    {\bf Data Set} & {\bf Metric} & {\bf BiGCN} & {\bf UPFD} & {\bf GLAN} & {\bf DUCK} & {\bf GAT} & {\bf RAW} & {\bf RA-SN}\\
    \midrule
    \multirow{4}{*}{{\bf Fakeddit}}
    & MF1 & 75.26$\pm$0.04 & 91.36$\pm$0.03 & \underline{91.69$\pm$0.02} &  91.62$\pm$0.02 & 91.54$\pm$0.03 & 90.76 & \textbf{92.43$\pm$0.04} \\
    & F1-T & 70.21$\pm$0.05 & 89.25$\pm$0.05 &\underline{89.72$\pm$0.03} & 89.61$\pm$0.02 &  89.52$\pm$0.03 & 88.61 & \textbf{90.60$\pm$0.05} \\
    & F1-R & 80.32$\pm$0.06 & 93.47$\pm$0.02 & \underline{93.67$\pm$0.03} & 93.62$\pm$0.02 & 93.57$\pm$0.04 & 92.92 & \textbf{94.26$\pm$0.06} \\
    & ACC & 76.30$\pm$0.05 & 91.88$\pm$0.03 & \underline{92.16$\pm$0.03} & 92.09$\pm$0.02 & 92.03$\pm$0.04 & 91.27 & \textbf{92.87$\pm$0.05} \\
    \midrule

    \multirow{4}{*}{{\bf PHEME}}
    & MF1 & 59.67$\pm$1.76 & \textbf{65.88$\pm$0.97} & 64.78$\pm$0.90 & 63.85$\pm$1.69 & \underline{65.34$\pm$1.74} & 61.77 & 64.17$\pm$1.99  \\
    & F1-T & 69.40$\pm$2.98 & \textbf{72.94$\pm$1.60} & \underline{71.55$\pm$2.98} & 70.51$\pm$0.92 & 71.42$\pm$2.96 & 69.47 & 71.47$\pm$2.77  \\
    & F1-R & 49.94$\pm$2.64 & \underline{58.81$\pm$2.39} & 58.01$\pm$3.33 & 57.18$\pm$3.53 & \textbf{59.26$\pm$3.15} & 54.07 & 56.88$\pm$2.76   \\
    & ACC & 62.14$\pm$2.13 & \textbf{67.42$\pm$0.91} & 66.34$\pm$1.24 & 65.16$\pm$1.16 & \underline{66.58$\pm$1.90} & 63.57 & 65.75$\pm$2.25 \\
    \midrule

    
    \multirow{4}{*}{{\bf RA-Gau}}
    & MF1
    & 62.94$\pm$1.41 & 65.47$\pm$0.83 & 67.43$\pm$0.35 & 67.11$\pm$0.56 & 63.80$\pm$1.31 & \underline{72.53}
    & \textbf{75.44$\pm$1.67} \\
    & F1-T
    & 55.90$\pm$1.49 & 56.85$\pm$1.24 & 59.24$\pm$1.69 & 58.79$\pm$2.23 & 54.26$\pm$2.52 & \underline{69.44}
    & \textbf{71.34$\pm$0.93} \\
    & F1-R & 69.99$\pm$4.20 & 74.08$\pm$1.82 & \underline{75.63$\pm$1.26} & 75.43$\pm$1.35 & 73.35$\pm$0.44 & 74.61
    & \textbf{79.54$\pm$2.44} \\
    & ACC & 64.48$\pm$2.25 & 67.66$\pm$1.29 & 69.54$\pm$0.59 & 69.28$\pm$0.54 & 66.34$\pm$0.81 & \underline{72.26}
    & \textbf{76.16$\pm$1.98} \\
    \midrule
    
    \multirow{4}{*}{{\bf RA-Uni}}
    & MF1 & 53.83$\pm$3.75 & 56.73$\pm$1.52 & 58.12$\pm$1.17 & 57.01$\pm$2.30 & 54.57$\pm$1.82 & \underline{72.07}
    & \textbf{74.91$\pm$0.75} \\
    & F1-T & 45.86$\pm$3.97 & 49.05$\pm$3.30 & 50.08$\pm$3.08 & 48.25$\pm$3.18 & 44.96$\pm$3.97 & \underline{67.57}
    & \textbf{70.18$\pm$0.64} \\
    & F1-R & 61.80$\pm$3.95 & 64.41$\pm$1.63 & 66.16$\pm$1.55 & 65.76$\pm$4.15 & 64.17$\pm$3.80 & \underline{74.56}
    & \textbf{79.63$\pm$1.33} \\
    & ACC & 55.24$\pm$3.88 & 58.16$\pm$1.21 & 59.74$\pm$0.79 & 58.96$\pm$3.00 & 56.80$\pm$2.28 & \underline{71.49}
    & \textbf{75.82$\pm$0.95} \\
    \midrule
    
    \multirow{4}{*}{{\bf RA-Zipf}}
    & MF1 & 70.49$\pm$1.45 & 71.54$\pm$1.45 & 73.53$\pm$0.73 & 74.82$\pm$0.88 & 66.35$\pm$1.17 & \underline{82.76}
    & \textbf{86.85$\pm$0.38} \\
    & F1-T & 65.69$\pm$1.02 & 64.29$\pm$1.98 & 66.92$\pm$1.63 & 68.96$\pm$1.40 & 56.39$\pm$2.32 & \underline{80.85}
    & \textbf{83.88$\pm$0.52} \\
    & F1-R & 75.29$\pm$3.08 & 78.79$\pm$1.60 & 80.14$\pm$0.24 & 80.69$\pm$0.68 & 76.32$\pm$1.08 & \underline{84.67} & \textbf{89.82$\pm$0.25} \\
    & ACC & 71.38$\pm$2.01 & 73.42$\pm$1.52 & 75.20$\pm$0.34 & 76.20$\pm$0.76 & 69.34$\pm$0.94 & \underline{82.97}
    & \textbf{87.52$\pm$0.34} \\

    \bottomrule
    \end{tabular}
    \caption{Rumor assertion performance. 
    The best results are highlighted in \textbf{bold}, and the second-best ones are \underline{underlined}.}
    \label{tab:1}
\end{table*}



\subsection{Overall Performance}
Table~\ref{tab:1} reports the performance of our RA-SN approach, compared with baselines, over real-world and synthetic graphs.
Overall, RA-SN achieves the best performance across most data sets (and comparable performance on one data set). On Fakeddit, RA-SN attains scores of 92.43\%, 90.60\%, and 94.26\% for MF1, F1-T, and F1-R, respectively. 
This shows that by explicitly modeling evidence from users' historical behaviors and probabilities, our RA-SN approach achieves excellent rumor assertion accuracy in large-scale, real-world social networks.


On PHEME data, RA-SN performs comparably to the best two, UPFD and GAT. This is because PHEME data contain sparse user interactions and limited historical behaviors, which restricts RA-SN from exploiting sufficient behavioral evidence. While our RA-SN does not directly model news content features, UPFD and GAT use additional information of content features, which increases the assertion accuracy. We would like to leave the interesting topic of combining content with user behaviors for hybrid rumor assertion as our future work.



Furthermore, compared to RAW, RA-SN improves by 2.4\%, 2.0\%, 2.8\%, and 2.2\% in MF1, F1-T, F1-R, and ACC, respectively. This indicates that behavior modeling and probability reconstruction improve performance on rumor assertion tasks.



For the three synthetic data sets RA-Gau, RA-Uni, and RA-Zipf, our RA-SN approach achieves MF1 scores of 75.44\%, 74.91\%, and 86.85\%, respectively. Overall, RA-SN has the best performance on all three synthetic graphs, confirming its robustness for different distributions.

\begin{table}[!t]
    \centering
    \resizebox{\linewidth}{!}{
    \begin{tabular}{ll|cccc}
    \toprule
    \textbf{Dataset} & \textbf{Metric}
    & \textbf{w/o EPR}
    & \textbf{w/o Index}
    & \textbf{w/o EPR \& Index}
    & \textbf{RA-SN} \\
    \midrule
    \multirow{4}{*}{\textbf{Fakeddit}}
    & MF1  & 91.07 & 91.83 & 90.76 & \textbf{92.43} \\
    & F1-T & 89.33 & 89.88 & 88.61 & \textbf{90.60} \\
    & F1-R & 92.93 & 93.78 & 92.92 & \textbf{94.26} \\
    & ACC  & 91.50 & 92.29 & 91.27 & \textbf{92.87} \\
    \midrule
    \multirow{4}{*}{\textbf{PHEME}}
    
    & MF1  & 62.02 & 63.11 & 61.77 & \textbf{64.17} \\
    & F1-T & 69.78 & 70.09 & 69.47 & \textbf{71.47} \\
    & F1-R & 53.92 & 53.22 & 54.07 & \textbf{56.88} \\
    & ACC  & 63.50 & 63.51 & 63.57 & \textbf{65.75} \\
    \bottomrule
    \end{tabular}
    }
    \caption{Performance of RA-SN variants on Fakeddit and PHEME.}
    \label{tab:ablation}
\end{table}

\subsection{Ablation Study}
Table~\ref{tab:ablation} presents the results of ablation studies on Fakeddit and PHEME to verify the effectiveness of the core modules within RA-SN.
Specifically, \textbf{w/o EPR} removes EPR, directly using raw historical user representations and index retrieval for rumor assertion; \textbf{w/o Index} removes the embedding-tree-based similar user retrieval and representation completion; and \textbf{w/o EPR \& Index} removes both EPR and index modules, relying instead directly on raw user statistics (i.e., RAW method) for rumor assertion.

The complete RA-SN model achieves the best performance on both Fakeddit and PHEME, validating the effectiveness of EPR and the embedding tree index. Removing EPR reduces the accuracy on both data sets: MF1 on PHEME drops from 64.17\% to 62.02\%, showing that probability reconstruction effectively improves user representation quality. Removing the embedding tree index also degraded the model performance, particularly on PHEME, where F1-R falls from 56.88\% to 53.22\%. This indicates that for users with insufficient historical evidence, the representation completion based on similar historical users can effectively mitigate the data sparsity. Moreover, our EPR component enhances user behavioral representations through probability reconstruction, while the embedding index addresses insufficient historical evidence for low-evidence users by retrieving similar users. Together, they improve our RA-SN query performance.


\begin{figure*}[t]
    \centering
    \subfigure[Threshold $\theta$]{
        \includegraphics[width=0.15\linewidth]{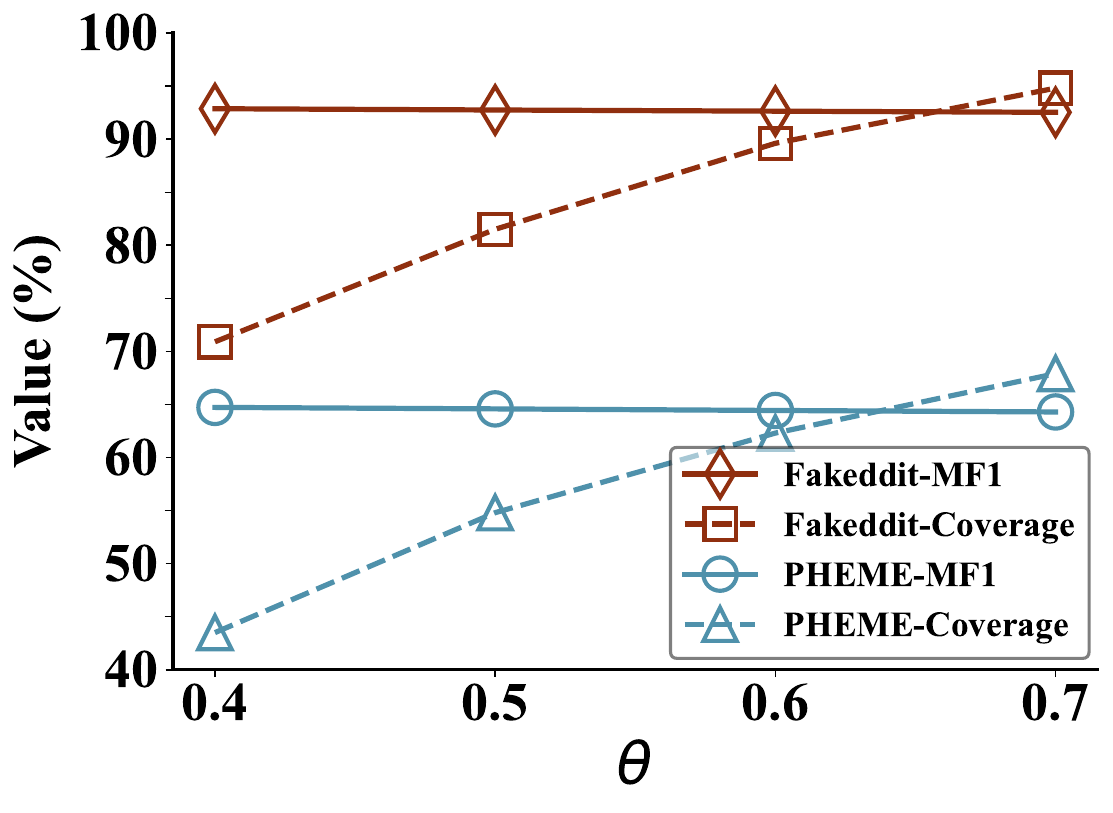}
        \label{subfig:parameter_theta}
    }\hfill
    \subfigure[Neighbors $K_g$]{
        \includegraphics[width=0.15\linewidth]{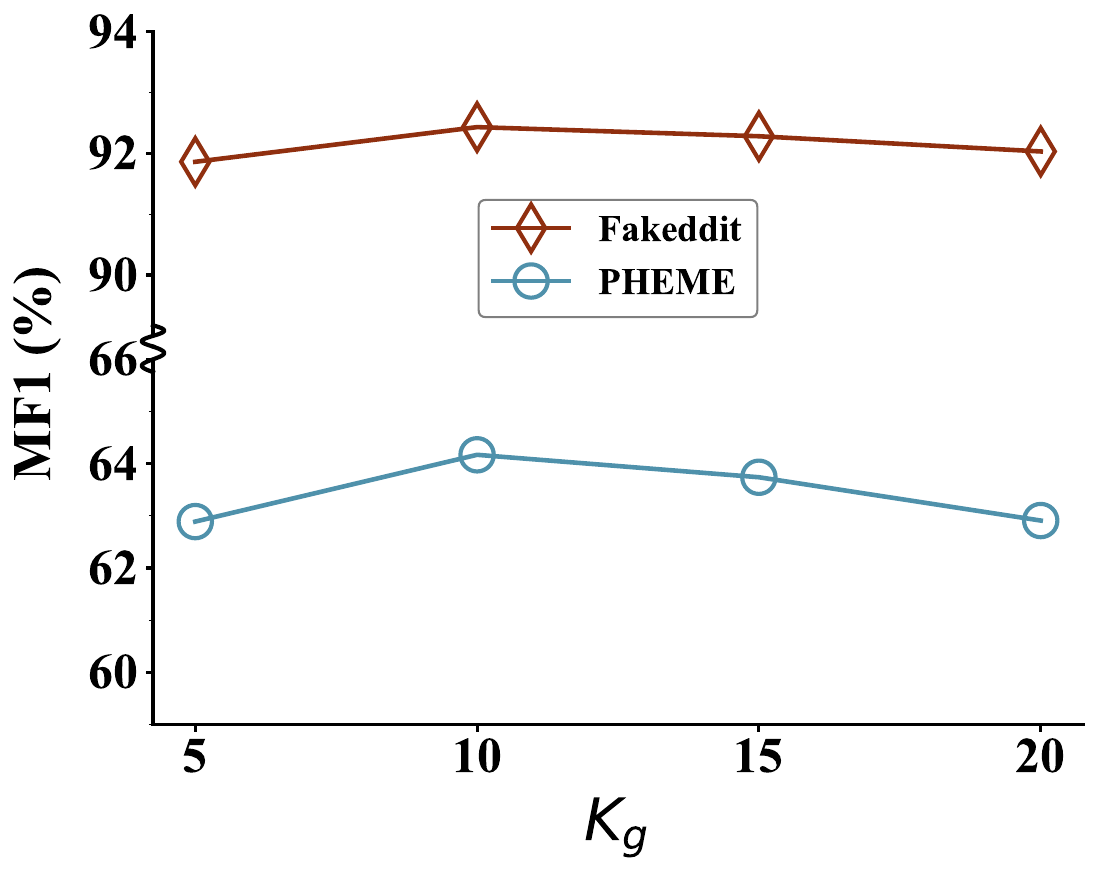}
        \label{subfig:KG}
    }\hfill
    \subfigure[Mask ratio $\rho$]{
        \includegraphics[width=0.15\linewidth]{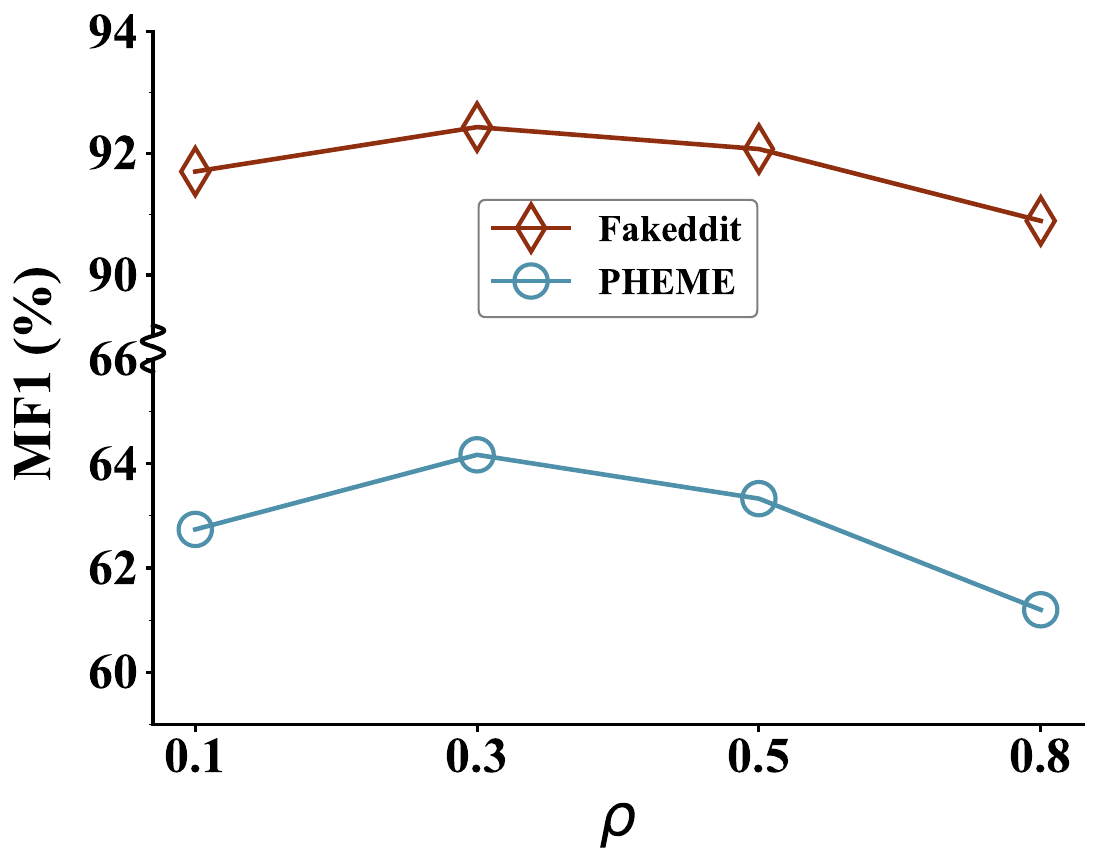}
        \label{subfig:rho}
    }\hfill
    \subfigure[Dimension $d$]{
        \includegraphics[width=0.15\linewidth]{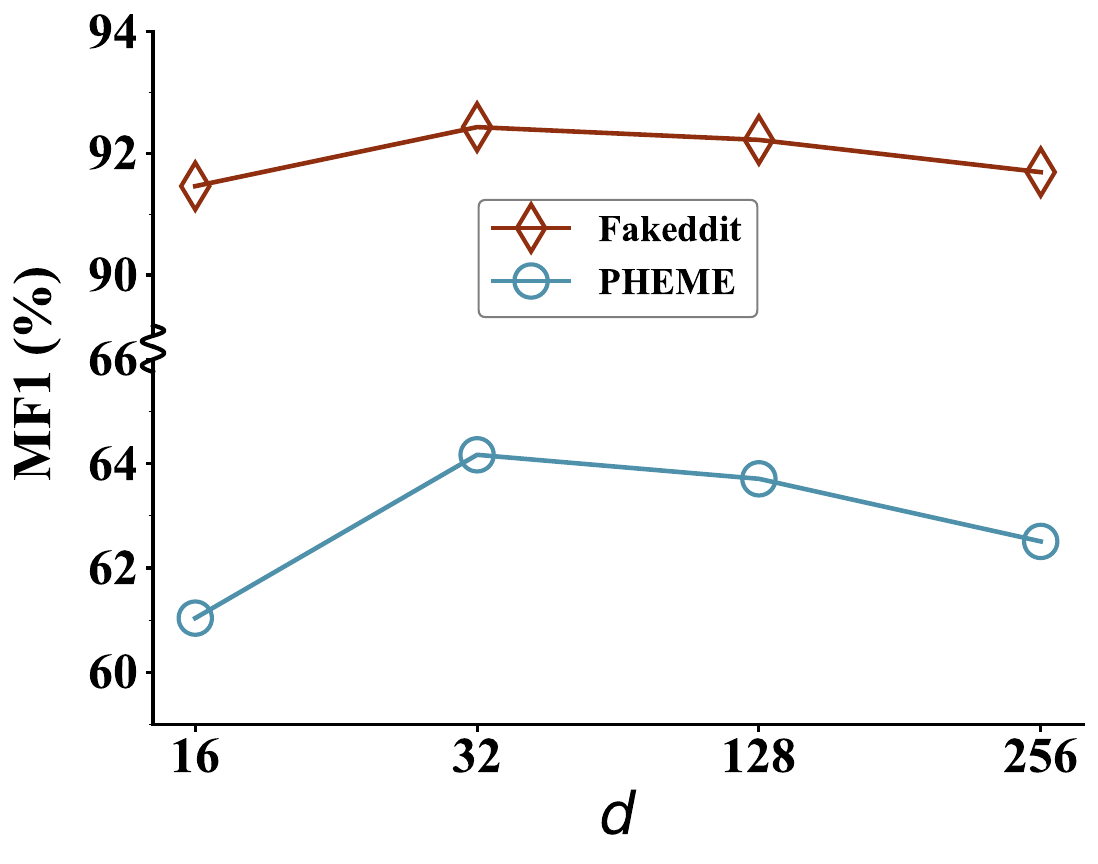}
        \label{subfig:D}
    }\hfill
    \subfigure[Retrieved size $K_r$]{
        \includegraphics[width=0.15\linewidth]{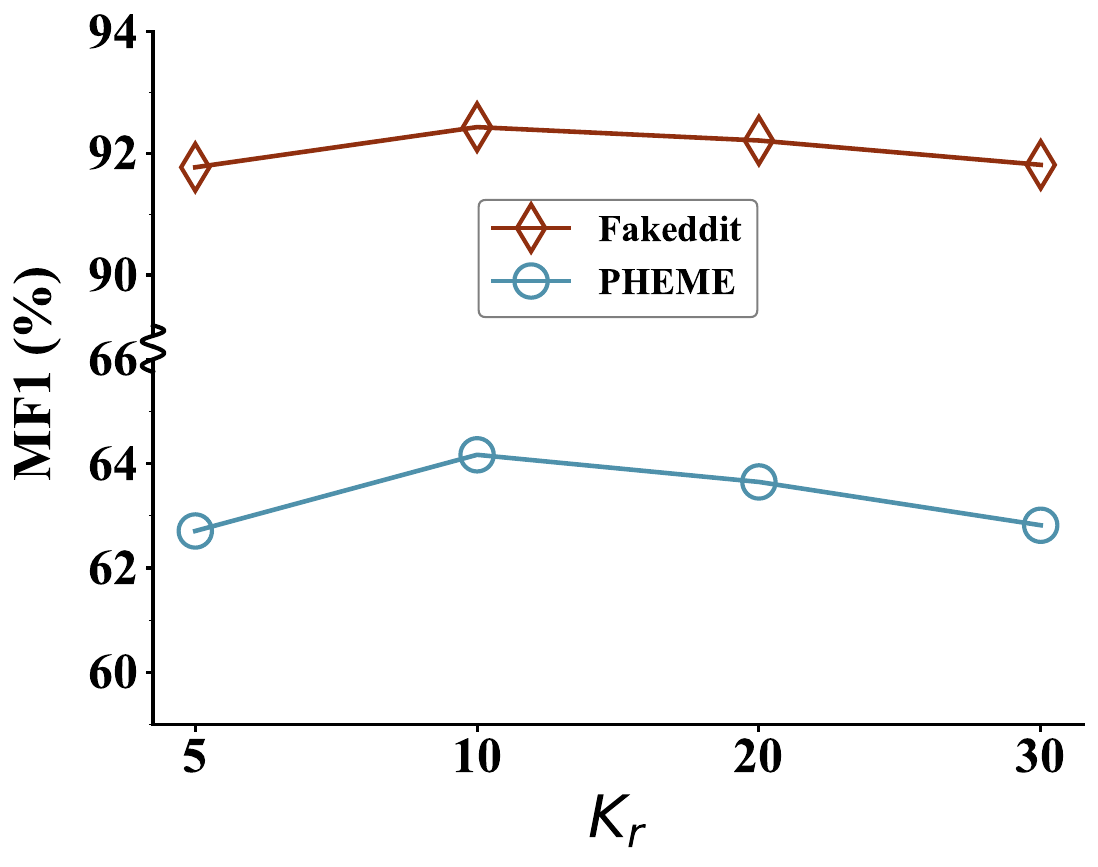}
        \label{subfig:KR}
    }\hfill
    \subfigure[Graph size $|V(G)|$]{
        \includegraphics[width=0.15\linewidth]{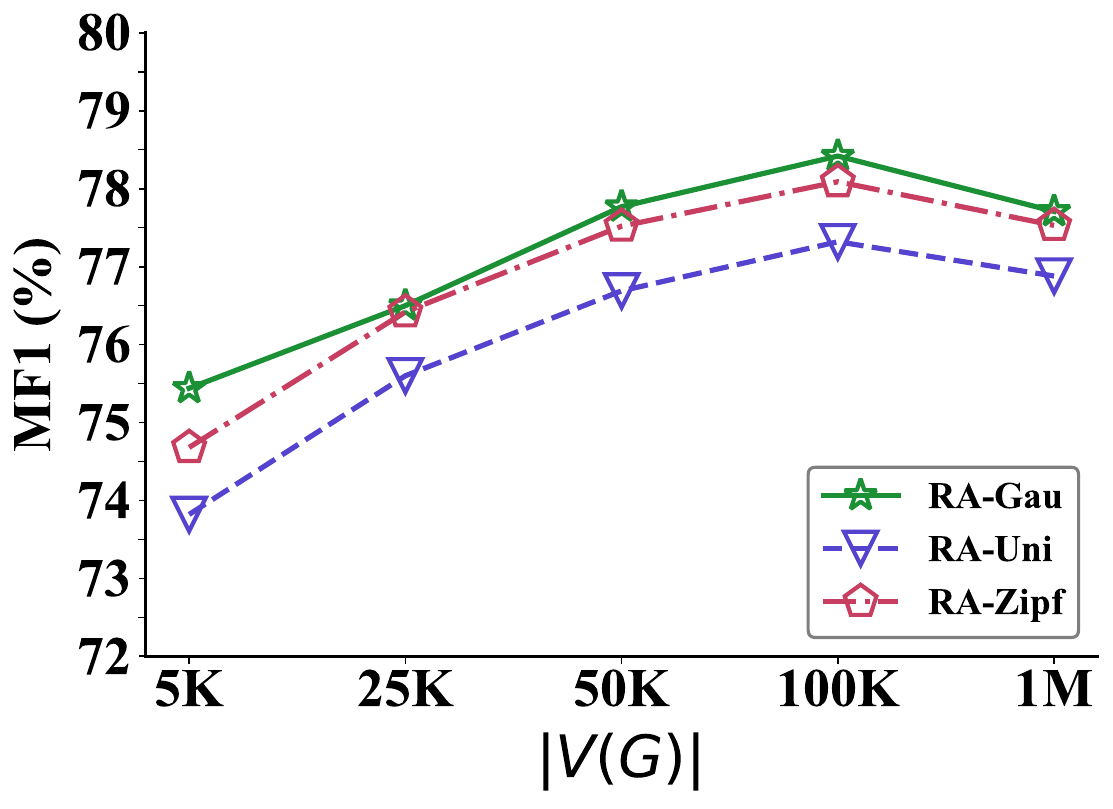}
        \label{subfig:vG}
    }
    \caption{Effectiveness evaluation w.r.t. parameters.}
    \label{fig:hyperparameter}
\end{figure*}

\begin{figure*}
    \centering
    \subfigure[index space cost]{
        \includegraphics[width=0.31\linewidth]{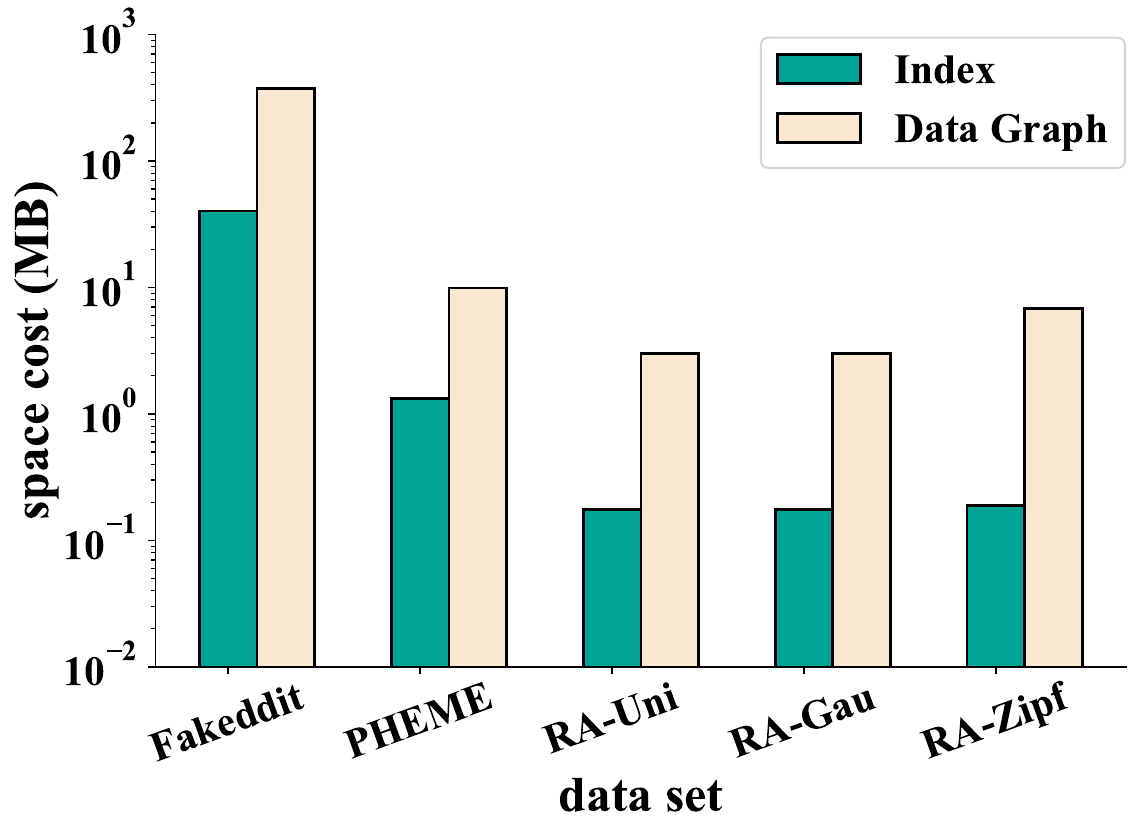}
            \label{subfig:index_space}
    } \hspace{-0.2cm}
    \subfigure[index query cost]{
        \includegraphics[width=0.31\linewidth]{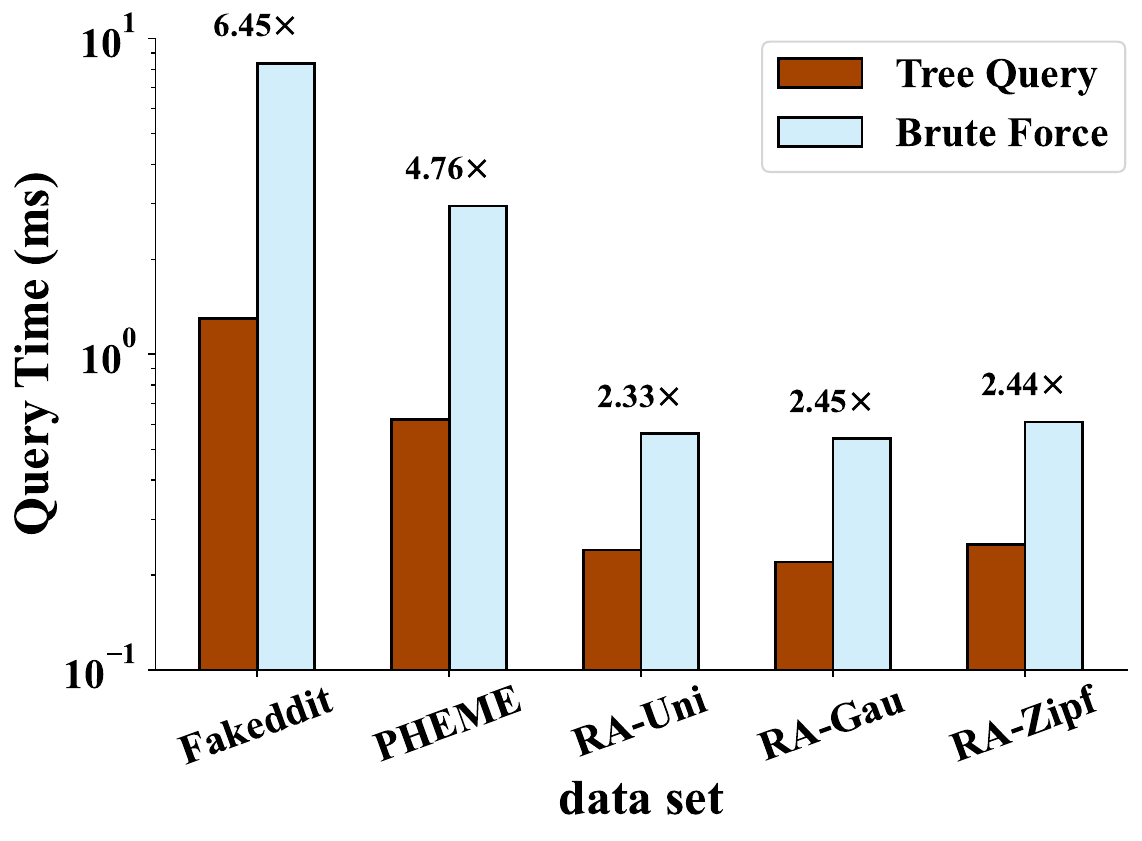}
        \label{subfig:index_query_time}
    }\hspace{-0.2cm}
    \subfigure[pruning power]{
        \includegraphics[width=0.31\linewidth]{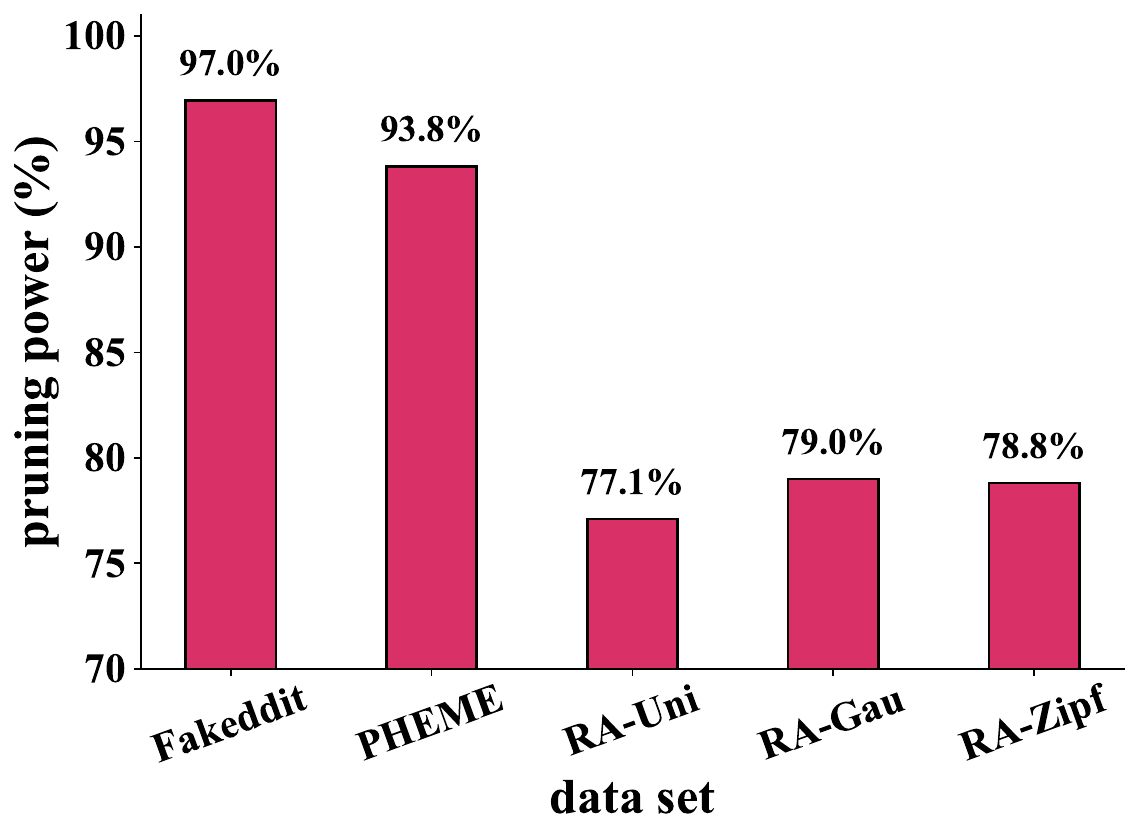}
        \label{subfig:pruning}
    }
    \caption{Space/time cost and pruning power evaluation of the indexing/pruning.}
    \label{fig:index}
\end{figure*}

\subsection{Effectiveness Evaluation w.r.t. Parameters}
Next, we test the robustness of our RA-SN approach for different parameters (e.g., $\theta$, $K_g$, $\rho$, $d$, $K_r$, and $|V(G)|$).


\noindent\textbf{The Effectiveness  w.r.t. the Entropy Threshold, $\theta$.} We use the \textbf{coverage} to represent the proportion of news items successfully verified. As shown in Figure~\ref{subfig:parameter_theta}, increasing the entropy threshold $\theta$ allows RA-SN to retain more uncertain news items, which improves the coverage. However, those additionally retained items typically have high entropy (i.e., low-confidence cases) and are difficult to classify correctly, leading to a gradual decrease in MF1. Nonetheless, for large-scale Fakeddit data, the overall MF1 remains above 92\%.



\noindent\textbf{The Effectiveness w.r.t. the \# of GNN Sampled Neighbors, $K_g$.} In Figure~\ref{subfig:KG}, when increasing $K_g$, the GNN provides sufficient local structural information while limiting noise propagation. However, large $K_g$ introduces more weakly correlated neighbors into message passing, leading to over-smoothing and accumulating behavioral noise. Consequently, the overall metrics initially rise and then decrease. The decrease is more noticeable for the PHEME data set because its smaller cascade scale makes the additionally sampled neighbors more likely to serve as weakly correlated or unstable evidence.

\noindent\textbf{The Effectiveness w.r.t. the Training Masking Ratio, $\rho$.} As the masking ratio increases in Figure~\ref{subfig:rho}, moderate evidence masking can enhance the reconstruction learning capability of EPR, but excessive masking (e.g., 0.8) leads to insufficient available context and neighbor recovery signals, thereby increasing the reconstruction error and prediction volatility. In contrast, PHEME, characterized by sparser historical interactions and behavior, is more sensitive to higher masking ratios.

\noindent\textbf{The Effectiveness w.r.t. the Embedding Dimension, $d$.} Figure~\ref{subfig:D} illustrates the results for embedding dimension $d$. When the embedding dimension $d$ increases, RA-SN can better encode user behaviors, neighbor contexts, and evidence reliability. However, higher dimensions tend to produce sparser representations and increase the risk of overfitting, thereby degrading performance. Nevertheless, MF1 scores on Fakeddit and PHEME remain above 91\% and 61\%, respectively, showing that RA-SN performs robustly across embedding dimensions.

\noindent\textbf{The Effectiveness w.r.t. the \# of Top Retrieved Nodes, $K_r$.} Figure~\ref{subfig:KR} reports the test results for $K_r$. As $K_r$ increases, more similar historical users can provide supplementary information for users with limited evidence, thereby improving the representation stability. However, a large $K_r$ introduces users with lower similarity and reliability, undermining the consistency of local behavioral patterns and causing the similar user completion process to degenerate into a simple neighborhood average. 
Therefore, a moderate $K_r$ value better balances sufficient information completion and the relevance of neighboring users.

\noindent\textbf{The Effectiveness w.r.t. the \# of Graph Size, $|V(G)|$.} We further evaluate the scalability of RA-SN on synthetic social-network graphs, ranging from $5K$ to $1M$ nodes. From Figure~\ref{subfig:vG}, as the graph size $|V(G)|$ increases, MF1 shows improvement. This is because a larger pool of historical users improves historical user coverage and reduces variance in probability estimates for users with limited evidence. 
When the graph size further increases (e.g., to $1M$), however, the newly added nodes consist mainly of long-tail users with limited evidence, while the scope of local message propagation remains constant. Consequently, larger scale does not yield sustained performance gains, and model performance gradually stabilizes or even slightly declines. These experimental results show that RA-SN benefits from larger-scale historical social-network graphs, while maintaining stable performance on ultra-large-scale graphs.


\subsection{Evaluation of the Indexing/Pruning Mechanisms}

Figure~\ref{subfig:index_space} examines the index space cost of two real-world and three synthetic graph data sets. The results show that our index is lightweight: compared to the original data graph, it achieves a maximum compression ratio of 9.3X for real-world data and 35.9X for synthetic graphs.  
Figure~\ref{subfig:index_query_time} presents the query time required for index queries over multiple data sets, compared to the brute-force search on the graph. From the figure, the tree index query takes only 1.8ms on the large-scale Fakeddit data, with a 6.45X improvement over the brute-force search.
Figure~\ref{subfig:pruning} illustrates the pruning power over real-world and synthetic graphs for our proposed pruning method in RA-SN. The pruning power reaches at least 93\% for real-world graphs, whereas that for synthetic graphs is above 77\%, which confirms the effectiveness (high pruning power) of our pruning method.



\section{Related Work}
\label{sec:related_work}

\vspace{1ex}\noindent{\bf Rumor Detection Over Social Networks.}
Rumor detection over social networks aims to determine the truthfulness of online information by analyzing news content, propagation patterns, and social context~\cite{bian2020rumor, shu2017fake, zubiaga2018detection}.
Previous research has primarily relied on text features, user information, and manually constructed statistical features~\cite{qazvinian2011rumor, vosoughi2015automatic, li2019rumor}.
Recently, researchers have increasingly adopted graph neural networks and language models to capture semantic features~\cite{bian2020rumor}. 
Existing work models the news dissemination process as a graph or tree and learns news representations based on these structures~\cite{bian2020rumor, yuan2019jointly}.
For example, BiGCN~\cite{bian2020rumor} models both top-down propagation and bottom-up diffusion with bidirectional graph convolution networks.
GLAN~\cite{yuan2019jointly} combines local semantic relations with a global heterogeneous graph involving source posts, reports, and users.
Although existing methods have achieved promising results, most focus on representation learning at the news or propagation cascade level and rely on consistent propagation structures during both training and inference. 
Furthermore, global social graph approaches tend to conflate user interactions across different news items, while news classifiers struggle to explicitly capture the differences in user behavior between real and fake news.
We separate the learning of past replies from evaluating new news. By modeling user behavior across different truthfulness contexts, we allow historical user replies to inform rumor assertion.

\vspace{1ex}\noindent{\bf Representation Learning for User Behavior.}
User behavior and social connections provide important clues for judging information authenticity. Existing studies typically characterize users based on historical posting and reposting preferences, interaction behaviors, and social connections, while employing graph representation learning to model dependencies between users~\cite{ shrestha2019learning, dou2021user, tian2022duck}.
For example, UPFD~\cite{dou2021user} integrates historical user preferences, news content, and interaction graph.
DUCK~\cite{tian2022duck} jointly models user networks and comment propagation networks to leverage social and conversational signals.
However, existing methods often represent users as low-dimensional vectors lacking explicit semantics or as static profiles, making it difficult to capture conditional user behaviors across varying veracity contexts. 
Sparse historical records can lead to unstable estimates of user credibility~\cite{hasan2025explainable, yang2024model}.
To address these challenges, we propose the Evidence-aware Probability Reconstructor (EPR), which characterizes conditional user behavior using a 6-dimensional representation with probabilistic semantics.
Through behavior message passing and evidence masking, EPR reconstructs behavioral probabilities from historical social structures, even when the evidence is otherwise insufficient.

\section{Conclusion}
In this paper, we propose a novel \textit{user behavior model} that captures users' ability to distinguish rumors on social networks, learned by our well-designed evidence-aware probability reconstructor. We formalize and tackle the problem of \textit{rumor assertion over social networks} (RA-SN), by devising a general framework for efficiently and effectively asserting whether news spread on social networks is a rumor via users' behaviors with high confidences. We conduct extensive experiments on real/synthetic social-network graphs to confirm the efficiency and effectiveness of our proposed RA-SN processing approach under various experimental settings.

\bibliography{aaai2026}
\end{document}